\documentclass[acmsmall]{acmart}
\usepackage{booktabs}
\usepackage{subcaption}
\usepackage{listings}
\usepackage{graphicx}
\usepackage{amsmath}
\usepackage{algpseudocode}
\usepackage{url}
\usepackage{siunitx}
\usepackage{xspace}
\usepackage{wrapfig}
\usepackage{mathtools}
\usepackage{adjustbox}

\DeclareMathOperator{\assign}{=}
\newcommand{\reduce}[1]{\mathrel{#1}=}
\newcommand{\iteration}[1]{\forall_{#1} \;}
\DeclareMathOperator{\where}{\textbf{where}}
\DeclareMathOperator{\seq}{; \;}

\definecolor{todocolor}{rgb}{0.8,0,0}
\definecolor{keywordcolor}{rgb}{0.5,0,0.5}

\newcommand{\figref}[1]{Figure~\ref{fig:#1}}

\newcommand{\secref}[1]{Section~\ref{sec:#1}}

\newcommand{\HIDE}[1]{}

\newcommand{\definitionspace}{\vspace{5px}}

\definecolor{textgray}{gray}{0.4}
\definecolor{mygray}{rgb}{0.5,0.5,0.5}
\newcommand\code[1]{\lstinline[columns=fullflexible, mathescape=true, basicstyle=\ttfamily]|#1|}

\newcommand{\tool}{\code{taco}\xspace}

\newcommand{\papertitle}{Sparse Tensor Algebra Optimizations with Workspaces}

\begin{document}

\title{\papertitle}

\author{Fredrik Kjolstad}
\email{fred@csail.mit.edu}

\author{Peter Ahrens}
\email{pahrens@csail.mit.edu}

\author{Shoaib Kamil}
\email{kamil@adobe.com}

\author{Saman Amarasinghe}
\email{saman@csail.mit.edu}

\begin{abstract}
This paper shows how to optimize sparse tensor algebraic expressions by
introducing temporary tensors, called workspaces, into the resulting loop
nests.  We develop a new intermediate language for tensor operations called
concrete index notation that extends tensor index notation.  Concrete index
notation expresses when and where sub-computations occur and what tensor they
are stored into.  We then describe the workspace optimization in this language,
and how to compile it to sparse code by building on prior work in the
literature.

We demonstrate the importance of the optimization on several important sparse
tensor kernels, including sparse matrix-matrix multiplication (SpMM), sparse
tensor addition (SpAdd), and the matricized tensor times Khatri-Rao product
(MTTKRP) used to factorize tensors.  Our results show improvements over prior
work on tensor algebra compilation and brings the performance of these kernels
on par with state-of-the-art hand-optimized implementations.  For example, SpMM
was not supported by prior tensor algebra compilers, the performance of MTTKRP
on the nell-2 data set improves by 35\%, and MTTKRP can for the first time have
sparse results.
\end{abstract}

\keywords{sparse tensor algebra, concrete index notation, optimization, 
temporaries}

\maketitle

\section{Introduction}
\label{sec:introduction}

% Temporary variables to optimize loops
Temporary variables are important for optimizing loops over dense tensors
(stored as arrays).  Temporary variables are cheaper to access than dense
tensors (stored as arrays) because they do not need address calculations, can
be kept in registers, and can be used to pre-compute loop-invariant
expressions.  Temporaries need not, however, be scalar but can also be
higher-order tensors called \textit{workspaces}.  Workspaces of lower dimension
(e.g., a vector) can be cheaper to access than higher-dimensional tensors
(e.g., a matrix) due to simpler address calculations and increased locality.
This makes them profitable in loops that repeatedly access a tensor slice, and
they can also be used to pre-compute loop-invariant tensor expressions.

% Temporary tensors to optimize sparse tensor loops
Temporary variables provide even greater opportunities to optimize loops that
compute operations on sparse tensors.  A sparse tensor's values are mostly
zeros and it can therefore be stored in a compressed data structure.  Dense
tensor temporaries can drastically reduce cost of access when they substitute
compressed tensors, as they have asymptotically cheaper random access and
insertion.  Random access and insertion into compressed tensors are
$\theta{}(\log{n})$ and $\theta{}(n)$ operations respectively as they require
search and data movement.  Furthermore, simultaneous iteration over compressed
data structures, common in sparse tensor codes, requires loops that merge
nonzeros using many conditionals.  By using dense tensor temporary variables,
of lower dimensionality to keep memory cost down, we can reduce cost of access,
insertion, and replace merge loops with random accesses.

% Importance of temporaries and lack of prior treatment
Prior work on sparse tensor compilation describes how to generate code for
sparse tensor algebra expressions~\cite{kjolstad2017}.  They do not, however,
consider temporary tensor workspaces nor do they describe optimizations that
use these.  Temporary tensor workspaces are an important tool in the
optimization of many sparse tensor kernels, such as tensor additions, sparse
matrix-matrix multiplication (SPMM)~\cite{gustavson1978}, and the matricized
tensor times Khatri-Rao product (MTTKRP)~\cite{smith2015}.  Without support for
adding workspaces we leave performance on the table.  In fact, the SpMM and
MTTKRP kernels are asymptotically slower without workspaces.

% Language and optimizations for temporaries
This paper presents an intermediate language, called concrete index notation,
that precisely describes when and where tensor sub-computations should occur
and the temporary variables they are stored in.  We then describe a compiler
optimization that rewrites concrete index notation to pre-compute
sub-expressions in workspace tensors, and a scheduling construct to request the
optimization.  This optimization improves the performance of sparse tensor code
by removing conditionals, hoisting loop-invariant sub-computations, and
avoiding insertion into sparse results.  Finally, we show how optimized
concrete index notation can be compiled to sparse code using the machinery
proposed by \citeauthor{kjolstad2017}~\shortcite{kjolstad2017}.
Our main contributions are:
\begin{description}

  \item[Concrete Index Notation] We introduce a new tensor expression
    representation that specifies loop order and temporary workspace variables.

  \item[Workspace Optimization] We describe a tensor algebra compiler
    optimization that removes expensive inserts into sparse results, eliminates
    merge code, and hoists loop invariant code.

  \item[Compilation] We show how to compile sparse tensor algebra expressions
    with workspaces, by lowering concrete index notation to the iteration
    graphs of~\citeauthor{kjolstad2017}~\shortcite{kjolstad2017}.

  \item[Case Studies] We show that the workspace optimization recreates several
    important algorithms with workspaces from the literature and generalizes to
    important new kernels.

\end{description}
We evaluate these contributions by showing that the performance of the
resulting sparse code is competitive with hand-optimized implementations with
workspaces in the MKL~\cite{mkl}, Eigen~\cite{eigen}, and
SPLATT~\cite{smith2015} high-performance libraries.

\section{Motivating Example}
\label{sec:example}

% Matrix multiplication example
We introduce sparse tensor data structures, sparse kernels, and the need for
workspaces with a sparse matrix multiplication kernel.  The ideas, however,
generalize to higher-order tensor kernels.  Matrix multiplication in linear
algebra notation is $A = BC$ and in tensor index notation it is $$A_{ij} =
\sum_k B_{ik} C_{kj}.$$

\begin{figure*}
  \begin{minipage}{0.49\linewidth}
    \begin{minipage}{\linewidth}
      \centering
      \includegraphics{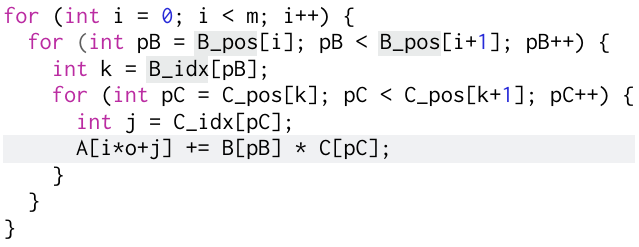}
      \vspace{-12px}
      \subcaption {
        $A_{ij} = \sum_k B_{ik} C_{kj}$ (sparse $B$, $C$)
      }
      \label{fig:matmul-sparse}
    \end{minipage}
    \begin{minipage}{\linewidth}
      \vspace{4px}
      \begin{minipage}{0.49\linewidth}
        \centering
        \includegraphics{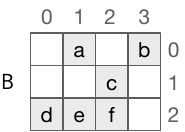}
        \subcaption {
          Dense $m \times o$ matrix $B$
        }
        \label{fig:matmul-matrix}
      \end{minipage}
      \begin{minipage}{0.49\linewidth}
        \centering
        \includegraphics{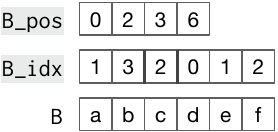}
        \subcaption {
          Sparse CSR index of $B$
        }
        \label{fig:matmul-index}
      \end{minipage}
    \end{minipage}
  \end{minipage}
  \begin{minipage}{0.49\linewidth}
    \centering
    \includegraphics{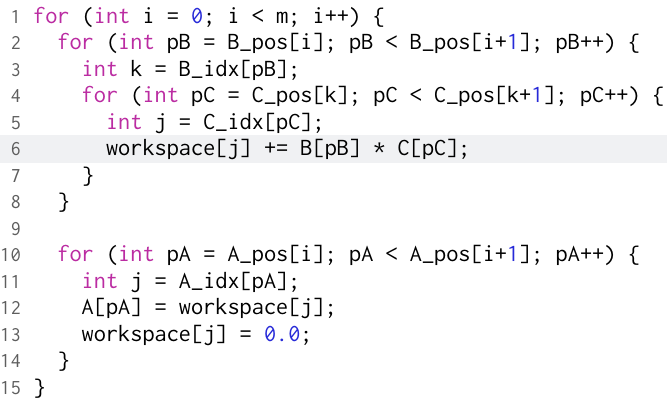}
    \subcaption {
      $A_{ij} = \sum_k B_{ik} C_{kj}$ (sparse $A$, $B$, $C$)
    }
    \label{fig:matmul-workspace}
  \end{minipage}
  \caption {\label{fig:matmul}
    Subfigures~a--c show a sparse matrix multiplication with a dense result,
    the matrix $B$, and its sparse CSR matrix data structure.  Subfigure~d shows
    the sparse multiplication after making the result also sparse.  Since the
    sparse matrix does not support fast random insert, we introduce a dense
    temporary workspace tensor.  The code to zero $A$ is
    omitted and result indices have been pre-assembled
    (\secref{workspace-assembly} discusses assembly). The code to allocate and
    initialize the workspace to zero has been ommited.
  }
\end{figure*}

% Matrix multiplication kernel
A matrix multiplication kernel's code depends on the storage formats of
operands and the result.  Many matrix storage formats have been proposed,
and can be classified as dense formats that store every matrix
component or sparse/compressed formats that store only the components that are
nonzero.  \figref{matmul} shows two matrix multiplication kernels using the
linear combination of rows algorithm.  We study this algorithm, instead of the
inner product algorithm, because its sparse variant has better asymptotic
complexity~\cite{gustavson1978} and because the inputs are all the same format
(row major).

% Sparse data structures and kernel
Sparse kernels are more complicated than dense kernels because they iterate
over sparse data structures.  \figref{matmul-sparse} shows a sparse matrix
multiplication kernel where the result matrix is stored dense row-major and the
operand matrices are stored using the compressed sparse row format
(CSR)~\cite{csr}.

% CSR layout
The CSR format and its column-major CSC sibling are ubiquitous in sparse linear
algebra libraries due to their generality and
performance~\cite{eigen,mkl,matlab}.  In the CSR format, each matrix row is
compressed (only nonzero components are stored).  This requires two index
arrays to describe the matrix coordinates and positions of the nonzeros.
\figref{matmul-matrix} shows a sparse matrix $B$ and \figref{matmul-index} its
compressed CSR data structure.  It consists of the index arrays \code{B_pos}
and \code{B_idx} and a value array \code{B}.  The array \code{B_idx} contains
the column coordinates of nonzero values in corresponding positions in
\code{B}.  The array \code{B_pos} stores the position of the first column
coordinate of each row in \code{B_idx}, as well as a sentinel with the number
of nonzeros ($\textrm{nnz}$) in the matrix.  Thus, contiguous values in
\code{B_pos} store the beginning and end [inclusive-exclusive) of a row in the
arrays \code{B_idx} and \code{B}.  For example, the column coordinates of the
third row are stored in \code{B_idx} at positions \mbox{$[$\code{B_pos[2]},
\code{B_pos[3]}$)$}.  Some libraries also stores the entries within each row in
order of ascending coordinate value, which results in better performance for
some algorithms.

% CSR iteration
Because matrix multiplication contains the sub-expression $B_{ik}$, the kernel
in~\figref{matmul-sparse} iterates over $B$'s sparse matrix data structure with
the loops over $i$ (line~1) and $k$ (lines~2--3).  The loop over $i$ is dense
because the CSR format stores every row, while the loop over $k$ is sparse
because each row is compressed.  To iterate over the column coordinates of the
$i$th row, the $k$ loop iterates over \mbox{$[$\code{B_pos[i]},
\code{B_pos[i+1]}$)$} in \code{B_idx}.  We have highlighted $B$'s index arrays
in \figref{matmul-sparse}.

% Workspace motivation
The kernel is further complicated when the result matrix $A$ is sparse, because
the assignment to $A$ (line~6) is nested inside the reduction loop $k$.  This
causes the inner loop $j$ to iterate over and insert into each row of $A$
several times.  Sparse data structures, however, do not support fast random
inserts (only appends).  Inserting into the middle of a CSR matrix costs
$\Theta{}(\textrm{nnz})$ because the new value must be inserted into the middle
of an array.  To get the $\Theta{}(1)$ insertion cost of dense formats, the
kernel in~\figref{matmul-workspace} introduces a dense workspace.  Such
workspaces and the accompanying loop transformations are the subject of this
paper.

% Workspace description
A workspace is a temporary tensor that is typically dense, with fast insertion
and random access.  Because values can be scattered efficiently into a dense
workspace, the loop nest $k,j$ (lines~2--8) in~\figref{matmul-workspace} looks
similar to the kernel in~\figref{matmul-sparse}.  Instead of assigning values
to the result matrix~$A$, however, it assigns them to a dense workspace vector.
When a row of the result is fully computed in the workspace, it is appended to
$A$ in a second loop over $j$ (lines~10--14).  This loop iterates over the row
in $A$'s sparse index structure, and thus assumes $A$'s CSR index has been
pre-assembled.  Pre-assembling index structures increases performance when
assembly can be moved out of inner loops and is common in material simulations.
\secref{workspace-assembly} describes the code to assemble result indices by
tracking the nonzero coordinates inserted into the workspace.

\section{Concrete Index Notation}
\label{sec:concrete-index-notation}

Specialized compilers for tensor and array operations succeed when they
appropriately simplify the space of operations they intend to compile.  For
this reason, many code generators and computational frameworks for tensor
algebra have adopted index notation as the input language to
optimize~\cite{kjolstad2017, vasilache2018, solomonik2014}.  Because index
notation describes what tensor algebra does but not how it is done, the user
does not mix optimization decisions with the algorithmic description.  It is
therefore easier to separately reason about different implementations, and the
algorithmic optimizations described in this work may be applied easily.  These
advantages come at the cost of restricting the space of operations that can be
described.

While index notation is good for describing the desired functionality, it
is unsuitable as an intermediate representation within a compiler because
it does not encode how the operation should be executed. There
are several existing representations one can use to fully describe how an index
expression might be computed, such as the code that implements the index
expression, sparse extensions of the polyhedral model~\cite{strout2012,
belaoucha2010}, or iteration graphs~\cite{kjolstad2017}. These representations,
however, are so general that it is difficult to determine when it is valid to
apply some of the optimizations described in this paper.

We propose a new intermediate language for tensor operations called concrete
index notation.  Concrete index notation extends index notation with constructs
that describe the way that an expression is computed.  In the compiler software
stack, concrete index notation is an intermediate representation between index
notation and the iteration graphs
of~\citeauthor{kjolstad2017}~\shortcite{kjolstad2017}.  A benefit of this
design is that we can reason about the legality of optimizations on the
concrete index notation without considering sparsity, which is handled by
iteration graphs lower in the stack.  We generate an expression in concrete
index notation as the first step in compiling a tensor expression in index
notation provided by the user.

Concrete index notation has three main statement types.  The assignment
statement assigns an expression result to a tensor element, the forall
statement executes a statement over a range inferred from tensor dimensions,
and the where statement creates temporaries that store subexpressions.

% Inner product matrix multiply
To give an example, let $A$, $B$, and $C$ be sparse matrices of dimension $I
\times J$, $I \times K$, and $K \times J$ where $A$ and $B$ are row-major (CSR)
and $C$ is column-major (CSC), and let $t$ be a scalar.  Consider the concrete
index expression for an \emph{inner products} matrix multiply, where each
element of $A$ is computed with a dot product of a corresponding row of $B$ and
column of $C$ (pseudo-code on right):

\begin{minipage}{\linewidth}
\vspace{1px}
\hspace{51.4mm}
\begin{minipage}{0.20\linewidth}
\center
$
  \iteration{ijk} A_{ij} \reduce{+} B_{ik} C_{kj}
$
\end{minipage}
\hspace{26mm}
\begin{minipage}{0.20\linewidth}
\center
\begin{lstlisting}[keywordstyle=\color{black}, numbers=none]
for (*$i \in I$*)
  for (*$j \in J$*)
    for (*$k \in K$*)
      (*$A_{ij} \reduce{+} B_{ik} * C_{kj}$*)
\end{lstlisting}
\end{minipage}
\vspace{1px}
\end{minipage}
The forall statements $\forall_{i}\forall_{j}\forall_{k}$, abbreviated as
$\forall_{ijk}$, specify the iteration order of the variables.  The resulting
loop nest computes in the inner $k$ loop the inner product of the $i$th row of
$B$ and the $j$th column of $C$.  The statement can be optimized by introducing
a scalar temporary $t$ to store the inner products as they are computed.  This optimization
can improve performance as it is cheaper to accumulate into a scalar due to
fewer address calculations.  The resulting concrete index notation adds a
$\where$ statement that introduces $t$ to hold intermediate computation of each
dot product:

\begin{minipage}{\linewidth}
\vspace{1px}
\hspace{38.0mm}
\begin{minipage}{0.20\linewidth}
\center
$
  \iteration{ij}
  \left( A_{ij} \assign t \right) \where
  \left( \iteration{k} t \reduce{+} B_{ik} C_{kj} \right)
$
\end{minipage}
\hspace{41.5mm}
\begin{minipage}{0.20\linewidth}
\center
\begin{lstlisting}[keywordstyle=\color{black}, numbers=none]
for (*$i \in I$*)
  for (*$j \in J$*)
    (*$t = 0$*)
    for (*$k \in K$*)
      (*$t \reduce{+} B_{ik} * C_{kj}$*)
    (*$A_{ij} \assign t$*)
\end{lstlisting}
\end{minipage}
\vspace{1px}
\end{minipage}

% Linear combinations of colums
The \emph{linear combinations of rows} matrix multiply computes rows of $A$ as
sums of the rows of $C$ scaled by rows of $B$.  When the matrices are sparse,
the linear combinations of rows matrix multiply is preferable to inner products
matrix multiply for two reasons.  First, sparse linear combinations of rows are
asymptotically faster because inner products must simultaneously iterate over
row/column pairs, which requires iterating over values that are nonzero in only
one matrix~\cite{gustavson1978}.  Second, linear combinations of rows work on
row-major matrices (CSR), while inner products require the second matrix to be
column-major (CSC).  It is often more convenient, as a practical matter, to
keep matrices ordered the same way.  We can express the linear combinations of
rows matrix multiply in concrete index notation by moving the $k$ loop above
the $j$ loop:

\begin{minipage}{\linewidth}
\vspace{1px}
\hspace{51.25mm}
\begin{minipage}{0.20\linewidth}
\center
$
  \iteration{ikj} A_{ij} \reduce{+} B_{ik} C_{kj}
$
\end{minipage}
\hspace{26.0mm}
\begin{minipage}{0.20\linewidth}
\center
\begin{lstlisting}[keywordstyle=\color{black}, numbers=none]
for (*$i \in I$*)
  for (*$k \in K$*)
    for (*$j \in J$*)
      (*$A_{ij} \reduce{+} B_{ik} * C_{kj}$*)
\end{lstlisting}
\end{minipage}
\vspace{1px}
\end{minipage}
This algorithm repeatedly computes and adds scaled rows to the matrix $A$.  If
$A$ is sparse, however, it is very expensive to repeatedly add rows.  We
therefore add a $\where$ statement that introduces a temporary vector $w$ to
hold the partial sums of rows:

\begin{minipage}{\linewidth}
\vspace{1px}
\hspace{33.6mm}
\begin{minipage}{0.20\linewidth}
\center
$
  \iteration{i}
    \left( \iteration{j} A_{ij} \assign w_j \right) \where
    \left( \iteration{kj} w_j \reduce{+} B_{ik} C_{kj} \right)
$
\end{minipage}
\hspace{43.8mm}
\begin{minipage}{0.20\linewidth}
\center
\begin{lstlisting}[keywordstyle=\color{black}, numbers=none]
for (*$i \in I$*)
  (*$w \assign 0$*)
  for (*$k \in K$*)
    for (*$j \in J$*)
      (*$w_j \reduce{+} B_{ik} * C_{kj}$*)
  for (*$j \in J$*)
    (*$A_{ij} \assign w_j$*)
\end{lstlisting}
\end{minipage}
\vspace{1px}
\end{minipage}
Note that the temporary $w$ is a vector, while the inner products temporary $t$
was a scalar.  The reason is that the we have added the loop $j$ underneath the
loop $k$ that we are reducing over.  The $j$ loop increases the distance
between the production of values on the right-hand-side of the $\where$ and
their consumption on the left hand side, and we must therefore increase the
dimensionality of the temporary by one to a vector of size equal to the range
of $j$.

\subsection{Definitions}

\begin{figure}
{\small
\newcommand\stmt{\text{statement}}
\newcommand\assignstmt{\text{assignment}}
\newcommand\forallstmt{\text{forall}}
\newcommand\wherestmt{\text{where}}
\newcommand\seqstmt{\text{sequence}}
\newcommand\mutatestmt{\text{mutation}}
\newcommand\expr{\text{expr}}
\newcommand\literalexpr{\text{literal}}
\newcommand\accessexpr{\text{access}}
\newcommand\indices{\text{indices}}
\newcommand\indexvar{\text{index}}
\newcommand\tensorvar{\text{tensor}}
\begin{minipage}[t]{0.45\linewidth}
\[\begin{array}{rll}
             \stmt := & \assignstmt  \\
                      & \forallstmt \\
                      & \wherestmt \\
                      & \seqstmt \\
       \assignstmt := & \accessexpr \, \assign \, \expr \\
                      & \accessexpr \, \reduce{+} \, \expr \\
                      & \dots \\
       \forallstmt := & \iteration{\indexvar} \; \stmt \\
        \wherestmt := & \stmt \; \where \; \stmt \\
          \seqstmt := & (\stmt \seq \stmt)*\\
\end{array}\]
\end{minipage}
\hspace{0.08\linewidth}
\begin{minipage}[t]{0.45\linewidth}
\[\begin{array}{rll}
       \accessexpr := & \tensorvar{}_{\indices} \\
          \indices := & \indexvar{}* \\
             \expr := & \literalexpr \\
                      & \accessexpr \\
                      & \mathbf{(} \expr {)} \\
                      & \expr + \expr \\
                      & \expr \, \expr \hspace{1cm} {\color{gray} // \, multiplication} \\
                      & \dots \\
\end{array}\]
\end{minipage}
  \caption { \label{fig:grammar}
    The grammar of concrete index notation.  The construct
    $\forall_{i\dots{}k}$ can be used as shorthand for
    $\forall_{i}\dots{}\forall_{k}$.
  }
}
\end{figure}

\figref{grammar} shows the grammar for concrete index notation.  Concrete index
notation uses \emph{index variables} to describe the iteration in tensor
algebra kernels.  Index variables are bound to integer values during execution,
and represent tensor coordinates in \emph{access expressions}. For an order $R$
tensor $A$ and distinct index variables $i_1, \dots, i_R$, the access
expression $A_{i_1\dots i_R}$ represents the single component of $A$ located at
coordinate $(i_1, ..., i_R)$. We sometimes abbreviate an access expression
$A_{i_1\dots i_R}$ as $A_{i\dots}$, and the sequence of index variables is
empty when we access a scalar tensor. A \emph{scalar expression} is defined to
be either a literal scalar, an access expression, or the result of applying a
binary operator to two scalar expressions, such as $A_{ij} \otimes 2$. Note
that binary operators are closed and are pure functions of their inputs.

Scalar expressions represent values, but statements modify the state of
programs. We refer to the state of a program in which a statement executes as
the environment, consisting of the names of tensors and index variables and the
values they hold. To retain some of the intuitive properties of index notation,
we restrict statements so that each modifies exactly one tensor.

The first concrete index notation statement we examine is the \emph{assignment
statement}, which modifies the value of a single tensor element.  Let
$A_{i\dots}$ be an access expression and let $E$ be a scalar expression of
variables in the unmodified environment.  The assignment statement $A_{i\dots}
\assign E$ assigns the value represented by $E$ to the element $A_{i\dots}$.
Although $E$ cannot contain the tensor $A$, assignment statements may use an
optional \emph{incrementing} form. For some binary operator $\oplus$, executing
$(A_{i\dots} \reduce{\oplus} E)$ assigns the value $(A_{i\dots} \oplus
E)$ to $A_{i\dots}$.

The \emph{forall statement} repeatedly binds an index variable to an integer
value. Let $S$ be a statement modifying the tensor $A$. We require that a
particular index variable $i$ in is only used to access modes with matching
dimension $D$, so if $i$ appears in $S$, then executing the forall statement
$\iteration{i} S$ executes $S$ once for each value of $i$ from $D$. Executing
$\forall_i S$ in the environment $V$ executes $S$ in a copy of $V$ where a new
binding $i$ has been added in a local scope, so changes to tensor $A \in V$ are
reflected in the original environment but the new binding for $i$ is not. To
avoid overwriting tensor values, we add a new constraint. If $\forall_i S$ is a
statement that modifies a tensor $A$ in an assignment statement $A_{j\dots}
\assign E$, then $i$ must be one of the index variables $j\dots$ which has not
yet been bound by $S$. We introduce \emph{multiple forall syntax} to simplify
writing multiple nested foralls in a row. Thus, $\iteration{i\dots}S$ is
equivalent to $\iteration{i_1}\iteration{i_2}... S$.

The \emph{where statement} precomputes a tensor subexpression. Let $S$ and $S'$
be statements which modify tensors $A$ and $A'$ respectively. The where
statement $(S \, \where \, S')$ then modifies the tensor $A$.  We execute the
where statement $(S \, \where \, S')$ in an environment $V$ in two steps.
First, we execute $S'$ in a copy of $V$ where $A$ has been removed.  Since our
statement may only modify $A$, $A'$ must not already be a variable in $V$ or
this expression would modify multiple tensors (we discuss the special case
where $A'$ and $A$ are the same tensor in the next paragraph). Next, we
execute $S$ in a copy of $V$ where $A'$ has been added. Note that this second
step does not add $A'$ to $V$, but changes to $A$ in this new environment are
reflected in $V$.

The \emph{sequence statement} modifies the same tensor multiple times in
sequence. The sequence statement $(S'; \, S)$ is like a where statement, except
the order of $S$ and $S'$ is swapped and instead of restricting $A'$ to be a
variable not in $V$, we say that $A'$ must be equal to $A$. Thus, the same
tensor is modified multiple times in a sequence. We may simplify multiple
nested sequence expressions in a row by omitting parenthesis so that $(S_0 \seq
S_1 \seq S_2 \seq \dots )$ is equivalent to $((S_0 \seq S_1) \seq S_{2}) \seq
\dots )$.

Finally, we describe when to initialize tensors. Notice that the
only two terminal statements in concrete index notation are the assignment
statement and the increment statement. Recall that each statement modifies
exactly one tensor. Before executing a concrete index statement that modifies a
tensor $A$ with an increment statement $A_{i\dots} \reduce{\oplus} E$, if $A$
is not defined in the environment then $A$ is initialized to the identity
element for the binary operation $\oplus$.

\subsection{Relationship to Index Notation}

Index notation is a compact notation for tensor operations that does not
specify how they are computed.  If $E$ is a scalar expression, the index
expression $A_{i\dots} = E$ evaluates $E$ for each value of $i\dots$ in the
dimensions of $A$ and sets $A_{i\dots}$ equal to the result. In this work, we
disallow the tensor $A$ from appearing in $E$.  We introduce a scalar
expression for index notation called the \emph{reduction expression}. The
reduction expression $\sum_{i\dots} E$ over the scalar expression $E$
evaluates to the sum of $E$ evaluated over the distinct values $i\dots$.

As an example, the following expression in index notation computes matrix
multiplication:
\[
  A_{ij} = \sum_{k}B_{ik}C_{kj}
\]

We can trivially convert an expression $A_{i\dots} = E_I$ in index notation to a
statement in concrete index notation $S_C$ as follows:
\begin{algorithmic}
\State Let $S_C$ be $A_{i\dots} = E_I$
\While{$S_C$ contains reduction nodes}
  \State Let $R = \sum_{j\dots} E_I$ be a reduction node in $S_C$.
  \State Replace $R$ with a fresh variable $t$ in $S_C'$
  \State Replace $S_C'$ with $S_C' \where (\iteration{j\dots} t \reduce{+} E_I)$
\EndWhile
\State Return $\iteration{i\dots} S_C$
\end{algorithmic}
The algorithm is improved if $R$ is always one of the outermost reduction nodes
in one of the the leftmost assignment statements $S_C'$ within $S_C$ that
contains a reduction expression.

\subsection{Reordering}

Reordering concrete index notation statements is useful for several reasons.
First, sparse tensors are sensitive to the order in which they are accessed.
For example, iterating over rows of a CSC matrix is costly. We can reorder
forall statements to yield better access patterns. We may also wish to reorder
to move loop-invariant where statements out of inner loops.  Critically, we may
need to reorder statements so that the preconditions for our workspace
optimization apply. When we reorder a concrete index statement, we want to
know that it will do the same thing as it used to.  We can express this
semantic equivalence by breaking down the transformation into small pieces.

We start by showing when we can rearrange forall statements.  Let $S$,
$\iteration{i}\iteration{j}S$, and $\iteration{j}\iteration{i}S$ be valid
statements in concrete index notation which do not contain sequence statements.
If $S$ modifies its tensor with an assignment statement or an increment
statement with an associative operator, then $\iteration{i}\iteration{j}S$ and
$\iteration{j}\iteration{i}S$ are semantically equivalent.

Next, we show when we can move a forall out of the left hand side of a where
statement.  Let $S_1$, $S_2$, $(\iteration{j} S_1) \where S_2$, and
$\iteration{j} (S_1 \where S_2)$ be concrete index statements which do not
contain sequence statements. If $S_2$ does not use the index variable $j$, then
$(\iteration{j} S_1) \where S_2$ and $\iteration{j} (S_1 \where S_2)$ are
semantically equivalent.

We can also move a forall out of both sides of a where statement.  Let $S_1$,
$S_2$, $(\iteration{j} S_1) \where (\iteration{j} S_2)$, and $\iteration{j}
(S_1 \where S_2)$ be concrete index statements which do not contain sequence
statements. If $S_2$ modifies its tensor with an assignment statement, then
$(\iteration{j} S_1) \where (\iteration{j} S_2)$ and $\iteration{j} (S_1 \where
S_2)$ are semantically equivalent.

Of course, we must rearrange nested where statements. We start by reordering
nests.  Let $S_1$, $S_2$, $S_3$, $(S_1 \where S_2) \where S_3$, and $S_1 \where
(S_2 \where S_3)$ be concrete index statements which do not contain sequence
statements. If $S_1$ does not use the tensor modified by $S_3$, then $(S_1
\where S_2) \where S_3$, and $S_1 \where (S_2 \where S_3)$ are semantically
equivalent.

We can also reorder right hand sides of $\where$ statements.  Let $S_1$, $S_2$,
$S_3$, $(S_1 \where S_2) \where S_3$, and $(S_1 \where S_3) \where S_2$ be
concrete index statements which do not contain sequence statements. If $S_2$
does not use the tensor modified by $S_3$ and $S_3$ does not use the tensor
modified by $S_2$, then $(S_1 \where S_2) \where S_3$, and $(S_1 \where S_3)
\where S_2$ are semantically equivalent.

\section{Workspace Optimization}
\label{sec:optimization}

\begin{figure}
  \begin{minipage}{0.460\linewidth}
    \begin{minipage}{\linewidth}
      \center
      {\small
      $
        \iteration{ij} a_{i} \reduce{+} B_{ij} C_{ij}
      $
      }
      \vspace{3mm}
    \end{minipage}
    \begin{minipage}{\linewidth}
      \center
      \includegraphics{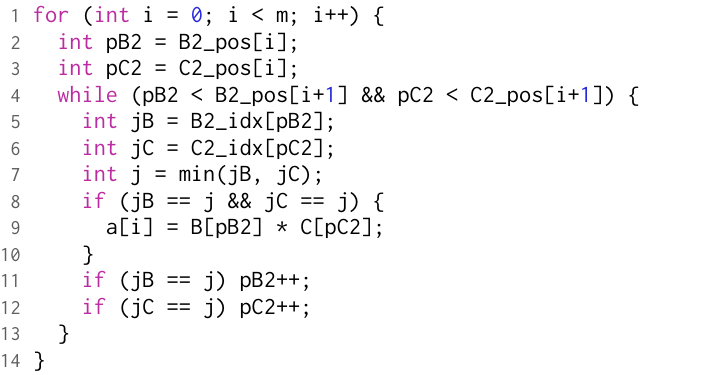}
    \end{minipage}
    \subcaption {\label{fig:examples-dotproducts-before}
      Before optimization the kernel iterates over the sparse intersection of
      each row of $B$ and $C$, by simultaneously iterating over their index
      structures to check if both have coordinates at each point.
    }
  \end{minipage}
  \hspace{0.02\linewidth}
  \begin{minipage}{0.508\linewidth}
    \begin{minipage}{\linewidth}
      \center
      {\small
      $
        \iteration{i}
        \left( \iteration{j} a_i \reduce{+} w_{j} C_{ij} \right) \where
        \left( \iteration{j} w_j \assign B_{ij} \right)
      $
      }
      \vspace{3mm}
    \end{minipage}
    \begin{minipage}{\linewidth}
      \center
      \includegraphics{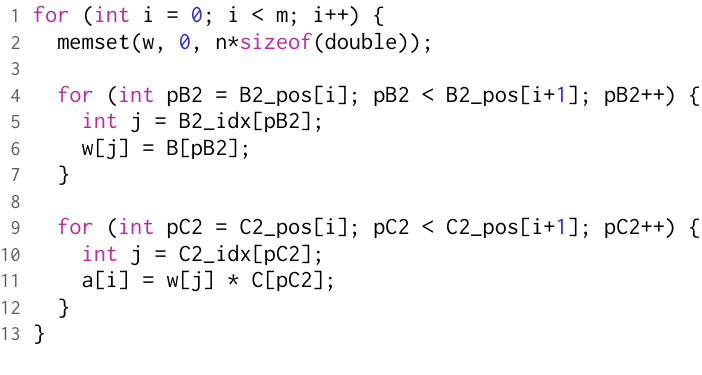}
    \end{minipage}
    \subcaption {\label{fig:examples-dotproducts-before}
      The workspace optimization introduces a $\where$ statement that results
      in two loops.  The first copies $B$ to a dense workspace $w$, and the
      second computes $A$ by iterating over $C$ and randomly accessing $w$.
    }
  \end{minipage}
  \vspace{-1.5mm}
  \caption {\label{fig:examples-dotproducts}
    Kernels that compute the sparse inner product of each pair of rows in the
    CSR matrices $B$ and $C$ \mbox{$a_{i} = \sum_{j} B_{ij} C_{ij}$} before and
    after applying the workspace optimization to the matrix $B$ over $j$.
    Sparse code generation is addressed in~\secref{compilation}.
  }
\end{figure}

% Introduction and motivation
The workspace optimization extracts and pre-computes tensor algebra
sub-expressions into a temporary workspace, using the concrete index notation's
$\where$ statement.  The workspace optimization can optimize sparse tensor
algebra kernels in the following three ways:
\begin{description}

  \item[Simplify merges]  Code to simultaneously iterate over multiple sparse
    tensors contains conditionals and loops that may be expensive.  By
    computing sub-expressions in dense workspaces, the code instead iterates
    over a sparse and dense operands (e.g.,~\figref{examples-dotproducts}).

  \item[Avoid expensive inserts]  Inserts into the middle of a sparse tensor,
    such as an increment inside of a loop, are expensive.  We can improve
    performance by computing the results in a workspace that supports fast
    inserts, such as a dense array or a hash map (e.g.,~\figref{case-spmm}).

  \item[Hoist loop invariant computations]  Computing a whole expression in the
    inner loop sometimes results in redundant computations.  Pre-computing a
    sub-expression in a separate loop and storing it in a workspace can hoist
    parts of a loop out of a parent loop (e.g.,~\figref{case-mttkrp-split}).
    
\end{description}
Many important sparse tensor algebra kernels benefit from the workspace
optimization, including sparse matrix multiplication, matrix addition, and the
matricized tensor times Khatri-Rao product.  In this section we describe the
optimization and give simple examples, and we will explore its application to
sophisticated real-world kernels in~\secref{case-studies}.

% The scheduling construct
To separate mechanism (how to apply it) and policy (whether to apply it), the
workspace optimization is programatically asked for using the \code{workspace}
method.  The method applies to an expression on the right-hand-side of an index
notation statement, and takes as arguments the index variables to apply the
workspace optimization to and a format that specifies whether the workspace
should be dense or sparse.  Implemented in C++, the API is 
\begin{lstlisting}[numbers=none]
void IndexExpr::workspace(std::vector<IndexVar> variables, Format format);
\end{lstlisting}
It can be used to generate the code in~\figref{matmul-workspace} as follows:
\begin{lstlisting}
Format CSR({dense, sparse});
TensorVar A(CSR), B(CSR), C(CSR);
IndexVar i, k, j;

IndexExpr mul = B(i,k) * C(k,j);
A(i,j) = sum(k)(mul);

mul.workspace({j}, Format(dense));
\end{lstlisting}
Lines 1--3 creates a CSR format, three tensor variables, and three index
variables to be used in the computation.  Lines 5--6 defines a sparse matrix
multiplication with index notation.  Finally, line 8 declares that the
multiplication should be pre-computed in a workspace, by splitting the \code{j}
loop into two \code{j} loops.  The optimization performs the following
transformation on the concrete index notation produced from the index notation:
$$
  \iteration{ikj} A_{ij} \reduce{+} B_{ik} C_{kj} \implies
  \iteration{i}
    \left( \iteration{j} A_{ij} \assign w_j \right) \where
    \left( \iteration{kj} w_j \reduce{+} B_{ik} C_{kj} \right),
$$
and result in the code shown on lines 2--9 in~\figref{matmul-workspace}.

\subsection{Definition and Preconditions}

% Intuition
The workspace optimization rewrites concrete index notation to pre-compute a
sub-expression.  The effect is that an assignment statement is split in two,
where one statement produces values for the other through a workspace.
\figref{examples-dotproducts} shows concrete index notation and kernels that
compute the inner product of each pair of rows from two matrices, before and
after the workspace optimization is applied to the matrix $B$ over $j$.  In
this example the optimization causes the while loop over $j$, that
simultaneously iterates over the two rows, to be replaced with a for loop that
independently iterates over each of the rows.  The for loops have fewer
conditionals, at the cost of reduced data locality.  Note that sparse code
generation is handled below the concrete index notation in the compiler stack,
as described in~\secref{compilation}.

% Definition
Let $(S, E, i\dots)$ be the inputs to the optimization, where $S$ is a
statement not containing sequences, $i\dots$ is a set of index variables, and
$E$ is an expression contained in an assignment or increment statement $S_A$
contained in $S$. If $S_A$ is the increment statement, let $\oplus$ be the
associated operator. The optimization rewrites the statement $S_A$ to
precompute $E$ in a workspace.  This operation may only be applied if every
operator on the right hand side of $S_A$ which contains $E$ distributes over
$\oplus$.
\begin{algorithmic}
\State Let $S_A'$ be $S_A$ where $E$ has been replaced by the access expression $w_{i\dots}$ where $w$ is a fresh tensor variable.
\State In $S$, replace $S_A$ with $S_A' \where (w_{i\dots} \reduce{\oplus} E)$.
\State Let $S_w$ be this where statement.
\While{$S_w$ is contained in a forall statement over an index variable $j$}
  \If{$j$ is used in both sides of $S_w$ and $j \in i\dots$}
    \State Move $\forall_j$ into both sides of $S_w$.
  \ElsIf{$j$ is used only in the left side of $S_w$}
    \State Move $\forall_j$ into the left side of $S_w$.
  \ElsIf{$j$ is used only in the right side of $S_w$}
    \State Move $\forall_j$ into the right side of $S_w$.
  \Else
    \State Stop.
  \EndIf
\EndWhile
\end{algorithmic}
The arrangement of the forall statements containing $S$ affects the results of
the optimization, so we may want to reorder before it is applied.  The order
(dimensionality) of the resulting workspace is the number of index variables in
$i\dots$ and the dimension sizes are equal to the ranges of those index
variables in the existing expression.

\subsection{Result Reuse}

% Motivation
When applying a workspace optimization $(S, E, i\dots)$ it sometimes pays to
use the left hand side of the assignment statement $S_A$ that contains $E$ as a
workspace.  Thus the expression $E$ is assigned to $w$ followed by $S_A$
rewritten as a incrementing assignment.  To support such mutation, we use the
sequence statement, which allows us to define a result and compute it in stages.
Result reuse is for example useful when applying the workspace optimization to
sparse vector addition with a dense result, as the partial results can be
efficiently accumulated into the result,
$$
\iteration{i} a_i \assign b_i + c_i \implies
\left( \iteration{i} a_i \assign b_i \seq \iteration{i} a_i \reduce{+} c_i
\right).
$$

% Application and precondition
The workspace optimization can reuse the result as a workspace if two
preconditions are satisfied.  The first precondition requires that the forall
statements on the two sides of the $\where$ statement are the same.  That is,
that the optimization does not hoist any computations out of a loop.  This
precondition ensures that the result does not get over-written by its use as a
workspace and is, for example, not satisfied by the second workspace
optimization to the MTTKRP kernel in~\secref{case-mttkrp}.  The second
precondition is that the expression $E$ is nested inside at most one operator
in $S_E$, which ensures we can rewrite the top expression to an incrementing
assignment.

\begin{figure}
  \begin{minipage}{0.460\linewidth}
    \begin{minipage}{\linewidth}
      \center
      {\small
        $
        \iteration{ij} A_{ij} \assign B_{ij} + C_{ij}
        $
      }
      \vspace{3mm}
    \end{minipage}
    \begin{minipage}{\linewidth}
      \center
      \includegraphics{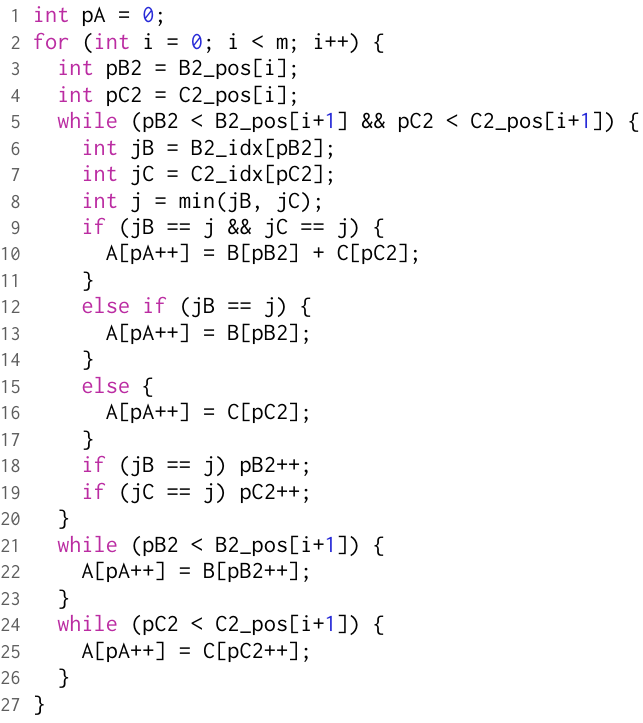}
    \end{minipage}
    \subcaption {\label{fig:examples-matadd-before}
      Before optimization the kernel iterates over the sparse union of each row
      of $B$ and $C$, by simultaneously iterating over their index structures
      to check if either have a coordinate at each point.
    }
  \end{minipage}
  \hspace{0.02\linewidth}
  \begin{minipage}{0.508\linewidth}
    \begin{minipage}{\linewidth}
      {\small
        \hspace{-4.1mm}
        $
        \iteration{i}
          \left( \iteration{j} A_{ij} \assign w_j \right) \where
            \left( \iteration{j} w_j \assign B_{ij} \seq
             \iteration{j} w_j \reduce{+} C_{ij} \right)
        $
      }
      \vspace{3mm}
    \end{minipage}
    \begin{minipage}{\linewidth}
      \center
      \includegraphics{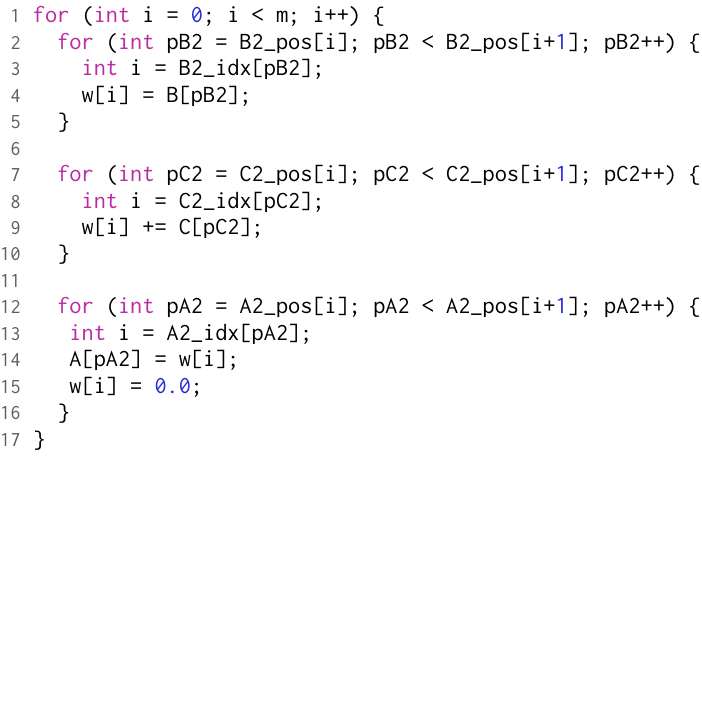}
    \end{minipage}
    \subcaption {\label{fig:examples-matadd-after}
      The workspace optimizations introduces a $\where$ statement with a
      sequence that results in three loops.  The first two adds $B$ and $C$ to
      $w$ and the second stores the results to $A$.
    }
  \end{minipage}
  \vspace{-1.5mm}
  \caption {\label{fig:examples-matadd}
    Sparse matrix addition $A_{ij} = B_{ij} + C_{ij}$ with CSR matrices before
    and after applying the workspace optimization twice over $j$.  The first
    application is to the $B_{ij}+C_{ij}$ expression, while the second
    application is to $B$ and reuses the workspace $w$ resulting in a sequence
    statement and the kernel above.
  }
\end{figure}

% Matrix Add Example
\figref{examples-matadd} shows a sparse matrix addition with CSR matrices
before and after applying the workspace optimization twice, resulting in a
kernel with three loops.  The first two loops add each of the operands $B$ and
$C$ to the workspace, and the third loop copies the non-zeros from the
workspace to the result $A$.  The first workspace optimization applies to the 
sub-expression $B_{ij}+C_{ij}$ over $j$ resulting in
$$
  \iteration{i}
    \left( \iteration{j} A_{ij} \assign w_j \right) \where
    \left( \iteration{j} w_j \assign B_{ij} + C_{ij} \right). 
$$
The second transformation applies to the $B_{ij}$ sub-expression on the
right-hand side of the $\where$.  Without result reuse the result would be
$$
  \iteration{i}
    \left( \iteration{j} A_{ij} \assign w_j \right) \where
    \left( \left( \iteration{j} w_j \assign v_j + C_{ij} \right) \where
           \left( \iteration{j} v_j \assign B_{ij} \right)
    \right),
$$
but with result reuse the two operands are added to the same workspace in a
sequence statement
$$
  \iteration{i}
    \left( \iteration{j} A_{ij} \assign w_j \right) \where
      \left( \iteration{j} w_j \assign B_{ij} \seq \iteration{j} w_j \reduce{+} C_{ij} \right).
$$

\subsection{Policy and Choice of Workspace}
\label{sec:tradeoffs}

% Costs of workspaces
The workspace optimization increases the performance of many important kernels
by removing inserts into sparse results, expensive merge code, and loop
invariant code.  It does, however, impose costs from constructing, maintaining,
and using workspaces.  Constructing a workspace requires a \code{malloc}
followed by a \code{memset} to zero its values and it must be reinitialized
between uses.  Furthermore, a workspace reduces temporal locality due to the
increased reuse distance from storing values to the workspace and later reading
them back to store to the result.

% Mechanism not policy
System designs are more flexible when they separate mechanism (what to do) from
policy (how to do it)~\cite{hansen70,wulf74}.  Performance is a key design
criteria in tensor algebra systems, so they should separate the policy
decisions of how to optimize code from the mechanisms that carry out the
optimization.  This paper focuses on optimization mechanisms.   

% Policy and scheduling language
We envision many fruitful policy approaches such as user-specified policy,
heuristics, mathematical optimization, machine learning, and autotuning.  We
leave the design of automated policy systems as future work. To facilitate
policy research, however, we have described an API for specifying workspace
optimizations.  We see this API as part of a scheduling language for index
notation.  The Halide system~\cite{ragan2012} has shown that a scheduling
language is effective at separating optimization mechanism from policy.
Scheduling languages leave users in control of performance, while freeing them
from low level code transformations.  The goal, of course, is a fully automated
system where users are freed from performance decisions as well.  Such a
system, however, also profits from a well-designed scheduling language, because
it it lets researchers explore different policy approaches without
re-implementing mechanisms.

% Other workspace choices
Furthermore, dense arrays are not the only choice for workspaces; a tensor of
any format will do.  The format, however, affects the generated code and its
performance.  The workspace optimization can be used to remove expensive
sparse-sparse merge code, and dense workspaces are attractive because they
result in cheaper sparse-dense merges.  An alternative is another format with
random access such as a hash map.  These result in slower
execution~\cite{patwary2015}, but only use memory proportional to the number
of nonzeros.

\section{Compilation}
\label{sec:compilation}

Concrete index notation is an effective intermediate representation for
describing important optimizations on index notation such as the workspace
optimization.  In this section we show how concrete index notation on sparse
and dense tensors is compiled to code.  We build on the work
of~\citeauthor{kjolstad2017}, which details a compiler for sparse tensor
expressions represented with an intermediate representation called iteration
graphs~\shortcite{kjolstad2017}.  We describe a process to convert concrete
index notation to iteration graphs that can then be compiled with their system.
We also show how their code generation machinery can be extended to assemble
workspace indices.

% Iteration Graph IR
The iteration graph intermediate representation for tensor algebra compilation
describes the constraints imposed by sparse tensors on the iteration space of
index variables~\cite{kjolstad2017}.  Sparse tensors provide an opportunity and
a challenge.  They store only nonzeros and loops therefore avoid iterating over
zeros, but they also enforce a particular iteration order because they encode
tensor coordinates hierarchically.

% Iteration graph definition and graphical notation
We construct iteration graph from concrete index notation, such as
\mbox{$\iteration{iklj} A_{ij} \reduce{+} B_{ilk}C_{lj}D_{kj}$} or
\mbox{$\iteration{ij} a_i \reduce{+} B_{ij}c_j+d_i$}.  In concrete index
notation, index variables range over the dimensions they index and computations
occur at each point in the iteration space.  Index variables are nodes in
iteration graphs and each tensor access, such as $B_{ij}$, becomes a path
through the index variables.

\begin{figure}
  \begin{minipage}[b]{0.22\linewidth}
    \centering
    \includegraphics{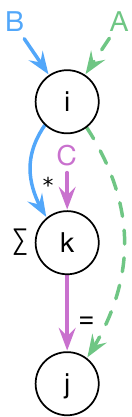}
    \vspace{2px}
    \subcaption {\label{fig:iteration-graphs-spmm}
      $A_{ij} = \sum_k B_{ik} C_{kj}$
    } 
  \end{minipage}
  \begin{minipage}[b]{0.22\linewidth}
    \centering
    \includegraphics{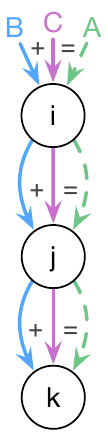}
    \subcaption {\label{fig:iteration-graphs-tadd}
      $A_{ijk} = B_{ijk} + C_{ijk}$
    } 
  \end{minipage}
  \begin{minipage}[b]{0.25\linewidth}
    \centering
    \includegraphics{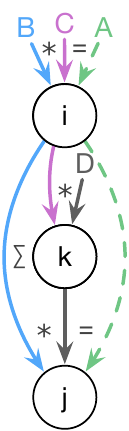}
    \subcaption {\label{fig:iteration-graphs-sddmm}
      $A_{ij} = \sum_k B_{ij} C_{ik} D_{kj}$
    }
  \end{minipage}
  \begin{minipage}[b]{0.26\linewidth}
    \centering
    \includegraphics{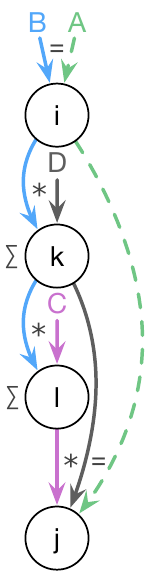}
    \subcaption {\label{fig:iteration-graphs-mttkrp}
      $A_{ij} = \sum_{lk} B_{ikl} C_{lj} D_{kj}$
    }
  \end{minipage}
  \caption {
    Four iteration graphs:
    (\subref{fig:iteration-graphs-spmm})~matrix multiplication,
    (\subref{fig:iteration-graphs-tadd})~tensor addition,
    (\subref{fig:iteration-graphs-sddmm}) sampled dense-dense matrix multiplication (SDDMM), and
    (\subref{fig:iteration-graphs-mttkrp}) matricized tensor times Khatri-Rao
    product (MTTKRP).
  }
  \label{fig:iteration-graphs}
\end{figure}

\begin{figure}
  \begin{minipage}[b]{0.39\linewidth}
    \centering
    \includegraphics{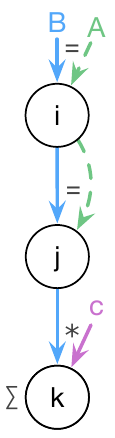}
    \subcaption {\label{fig:iteration-graphs-ttv}
      Iteration Graph
    }
  \end{minipage}
  \begin{minipage}[b]{0.60\linewidth}
    \includegraphics{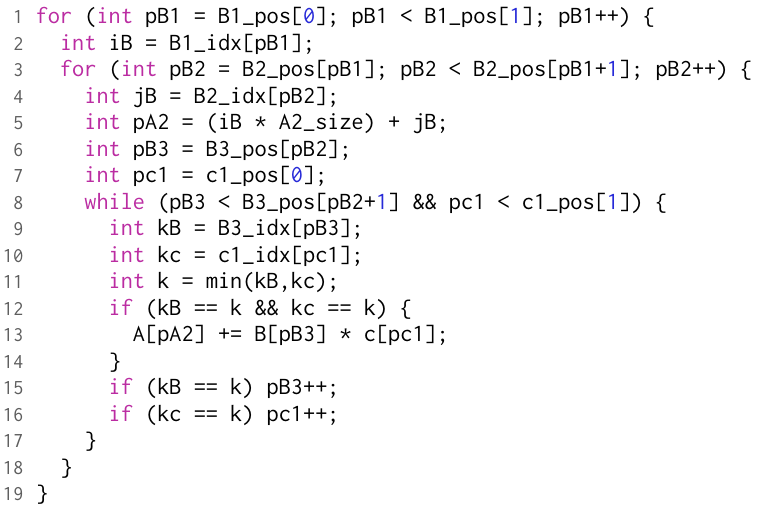}
    \subcaption {\label{fig:code-ttv}
      Generated code when $A$ is dense while $B$ and $c$ are sparse. 
    }
  \end{minipage}
  \caption {\label{fig:ttv}
    Tensor-vector multiplication $A_{ij} = \sum_k B_{ijk} c_{k}$.  Each index
    variable becomes a loop over the sparse tensor indices of its incoming
    paths.  The $k$ loop iterates over the intersection of the last dimension
    of $B$ and $c$.
  }
\end{figure}

% Matrix-vector multiplication example
\figref{iteration-graphs} shows several iteration graphs, including matrix
multiplication, sampled dense-dense matrix multiplication from machine
learning~\cite{sddmm}, and the matricized tensor times Khatri-Rao product used
to factorize tensors~\cite{bader2007}.  The concrete index notation for
tensor-vector multiplication is, for example, $$\iteration{ijk} A_{ij}
\reduce{+} B_{ijk} c_{k}.$$ The corresponding iteration graph in
\figref{iteration-graphs-ttv} has a node for each index variable $i$, $j$, and
$k$ and a a path for each of the three tensor accesses $B_{ijk}$ (blue),
$c_{k}$ (purple), and $A_{ij}$ (stippled green).  We draw stippled paths for
results.  \figref{code-ttv} shows code generated from this iteration graph when
$B$ and $c$ are sparse.  Each index variable node becomes a loop that iterates
over the sparse tensor indices belonging to the incoming edges.

% Sparse iteration, merging, intersections, unions, sparse-dense and workspaces
Two or more input paths meet at an index variable when it is used to index into
two or more tensors.  The iteration space of the tensor dimensions the variable
indexes must be merged in the generated code.  The index variables are
annotated with operators that tell the code generator what kind of merge code
to generate.  If the tensors are multiplied then the generated code iterates
over the intersection of the indexed tensor dimensions
(\figref{examples-dotproducts}).  If they are added then it iterates over their
union (\figref{examples-matadd}).  If more than two tensors are indexed by the
same index variable, then code is generated to iterate over a mix of
intersections and unions of tensor dimensions.

% Constructing iteration graphs
Iteration graphs are a hierarchy of index variable nodes, together with tensor
paths that describe tensor accesses.  Constructing an iteration graph from
concrete index notation is a two-step process:
\begin{description}

  \item[Construct Index Variable Hierarchy] To construct the index variable
    hierarchy, traverse the concrete index notation.  If the forall statement
    of an index variable $j$ is nested inside the forall statement of index
    variable $i$, then we also place $j$ under $i$ in the iteration graph.
    Furthermore, the index variables of two forall statements on different
    sides of a $\where$ statement become siblings.

  \item[Add Tensor Paths] To add the paths, visit each tensor access
    expression.  For each access expression, add a path between the index
    variables used to access the tensor.  The order of the path is determined
    from the order the dimensions are stored.  If, for example, the access
    expression is $B_{ij}$ then add the path $(i,j)$ if the matrix is row major
    (e.g., CSR) and the path $(j,i)$ if the matrix is column major (e.g., CSC).

\end{description}

% Code generation
\citeauthor{kjolstad2017} described an algorithm to generate code from
iteration graphs, including a mechanism called merge lattices to generate code
to co-iterate over tensor dimensions~\shortcite{kjolstad2017}.  Understanding
our workspace optimization does not require understanding the details of the
code generation algorithm or merge lattices.  We should note, however, that the
performance of code that merges sparse tensors may suffer from many
conditionals.  Code to co-iterate over a combination of a single sparse and one
or more dense tensors, on the other hand, does not require conditionals.  One
of the benefits of introducing a workspace is to improve performance by turning
sparse-sparse iteration into sparse-dense iteration.

\label{sec:workspace-assembly}

\begin{figure}
    \includegraphics{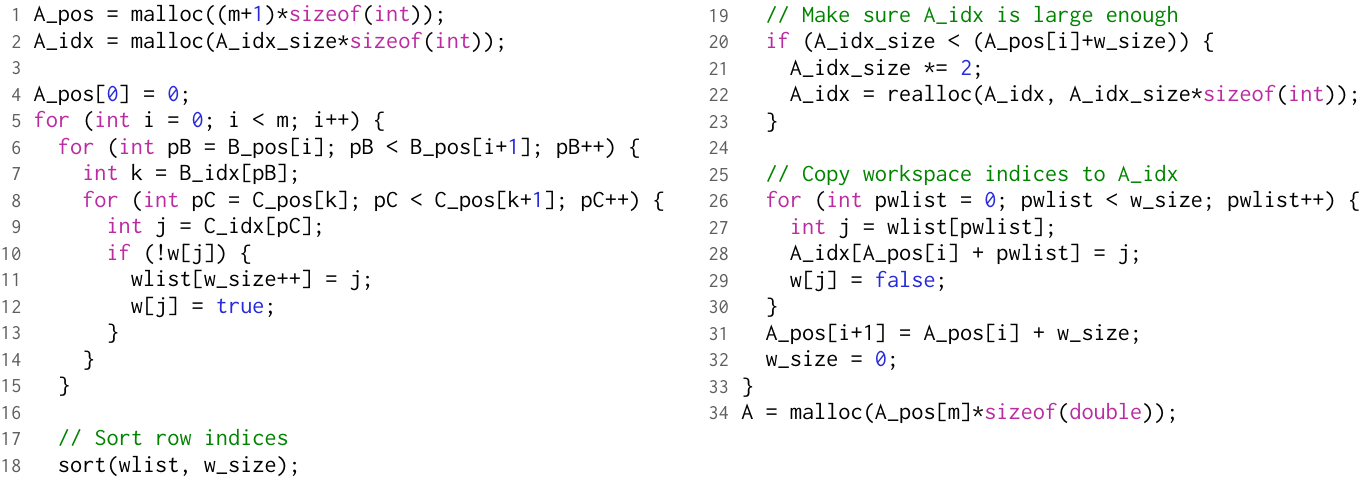}
  \caption {
    Sparse matrix multiply assembly kernel (the compute kernel is
    in~\figref{matmul-workspace}).  The coordinates of row $i$ are inserted
    into \code{wlist} on line~11 and copied to \code{A} on line 28.  The array
    \code{w} guards against redundant inserts.
  }
  \label{fig:code-matmul-assembly}
\end{figure}

% Numeric/symbolic separation
In code listings that compute sparse results, we have so far shown only kernels
that compute results without assembling sparse index structures
(Figures~\ref{fig:matmul-workspace}, \ref{fig:examples-matadd-after},
and~\ref{fig:code-mttkrp-splits}).  This let us focus on the loop structures
without the added complexity of workspace assembly.  Moreover, it is common in
numerical code to separate the kernel that assembles index structures (often
called symbolic computation) from the kernel that computes values (numeric
computation)~\cite{gustavson1978, heath1991}.  The code generation algorithm
for iteration graphs can emit either, or a kernel that simultaneously assembles
the result index structures and computes its values.

% Workspace consideration
When generating assembly kernels from iteration graphs, a workspace consists of
two arrays that together track its nonzero index structure.  The first array
\code{wlist} is a list of coordinates that have been inserted into the
workspace, and the second array (\code{w}) is a boolean array that guards
against redundant inserts into the coordinate list.

% Matrix multiplication example
\figref{code-matmul-assembly} shows assembly code for sparse matrix
multiplication generated from the iteration graph in~\figref{case-spmm-split}.
It is generated from the same iteration graph as the compute kernel
in~\figref{matmul-workspace}, so the loop structure is the same except for the
loop to copy the workspace to \code{A} on line~26.  In compute kernels, the
index structure of \code{A} must be pre-assembled, so the code generation
algorithm emits a loop to iterate over \code{A}.  In an assembly kernel,
however, it emits code to iterate over the index structure of the workspace.
Furthermore, the assembly kernel inserts into the workspace index
(\code{wlist}), on lines~10--13, instead of computing a result, and sorts the
index list on line~18 so that the new row of \code{A} is ordered.  Note that
the sort is optional and only needed if the result must be ordered.  Finally,
the assembly kernel allocates memory on lines~1--2, 20--23 (by repeated
doubling), and 34.

\section{Case Studies}
\label{sec:case-studies}

In this section we study three important linear and tensor algebra
expressions that can be optimized with the workspace optimization.  The
resulting kernels are competitive with hand-optimized kernels from the
literature~\cite{gustavson1978,smith2015,eigen}.  The optimization, however,
generalizes to an uncountable number of kernels that have not been implemented
before.  We will show one example, MTTKRP with sparse matrices,
in~\secref{case-mttkrp}.

\subsection{Matrix Multiplication}
\label{sec:case-spmm}

\begin{figure}
  \centering
  \begin{minipage}{0.49\linewidth}
    \center
    {\small
      $\iteration{ikj} A_{ij} \reduce{+} B_{ik} C_{kj}$
    }\\
    \includegraphics{case-spmm-original}
    \subcaption{
      Iteration graph before workspace optimization
    }
    \label{fig:case-spmm-original}
  \end{minipage}
  \begin{minipage}{0.49\linewidth}
    \center
    {\small
      $\iteration{i}
        ( \iteration{j} A_{ij} \assign w_j ) \, \where \;
        ( \iteration{kj} w_j \reduce{+} B_{ik} C_{kj} )
      $
    }\\
    \includegraphics{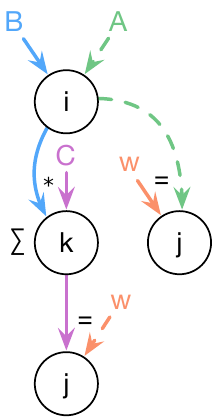}
    \subcaption{
      Iteration graph after workspace optimization
    }
    \label{fig:case-spmm-split}
  \end{minipage}
  \caption { \label{fig:case-spmm}
    Matrix multiplication \mbox{$A_{ij} = \sum_k B_{ik} C_{kj}$} using the
    linear combination of rows algorithm with all matrices in the CSR format.
    Pre-computing row-computations in a workspace recreates
    \citeauthor{gustavson1978}'s algorithm~\shortcite{gustavson1978} shown
    showed~\figref{matmul-workspace}.
  }
\end{figure}

% Linear combination sparse matrix multiplication
The preferred algorithm for multiplying two sparse matrices is to compute the
linear combinations of rows or columns~\cite{eigen,csparse,matlab,julia}.  This
algorithm was introduced by
\citeauthor{gustavson1978}~\shortcite{gustavson1978}, who showed that it is
asymptotically superior to computing inner products when the matrices are
sparse.  Furthermore, both operands and the result are the same format.  A
sparse inner product algorithm inconveniently needs the first operand to be row
major (CSR) and the second column major (CSC).

% Algorithm before split
\figref{case-spmm-original} shows the concrete index notation and iteration
graph for a linear combination of rows algorithm, where the matrices are stored
in the CSR format.  The iteration graph shows an issue at index variable $j$.
Because the assignment to $A$ at $j$ is dominated by the summation index
variable $k$ in the iteration graph, the generated code must repeatedly add new
values into $A$.  This is expensive when $A$ is sparse due to costly inserts
into its sparse data structure.

% Algorithm after split
In \figref{case-spmm-split}, the concrete index notation has been optimized to
pre-compute the $B_{ik} C_{kj}$ sub-expression in a workspace.  In the
resulting iteration graph this results in $j$ being split into two new index
variables.  The first accumulates values into a dense workspace $w$, while
$j_A$ copies the nonzero values from the workspace to $A$.  Because the
workspace is dense, the merge with $C$ at $j_C$ is trivial: the kernel iterates
over $C$ and scatters values into $w$.  Furthermore, the second index variable
$j_A$ is not dominated by the summation variable $k$ and values are therefore
appended to $A$.

% Kernel code
The code listing in \figref{matmul-workspace} showed the code generated from a
matrix multiplication iteration graph where the assignment operator has been
split.  Each index variable results in a loop, loops generated from index
variables connected by an arrow are nested, and loops generated from index
variables that share a direct predecessor are sequenced.  The last $j$ loop
copies values from the workspace to $A$, so it can either iterate over the
nonzeros of the workspace or the index structure of $A$.  The loop on
lines~10--14 in the code listing iterates over the index structure of $A$,
meaning it must be pre-assembled before this code is executed.  The alternative
is to emit code that tracks the nonzeros inserted into the workspace, but this
is more expensive. It is sometimes more efficient to separate the code that
assembles $A$'s index structure from the code that computes its
values~\cite{gustavson1978}.  We discussed code generation for pure assembly
and fused assembly-and-compute kernels in \secref{workspace-assembly}.  These
kernels cannot assume the results have been pre-assembled and must maintain and
iterate over a workspace index.

\subsection{Matrix Addition}
\label{sec:case-spadd}

\begin{figure}
  \centering
  \begin{minipage}{0.40\linewidth}
    \center
    {
      $\iteration{ij} A_{ij} = B_{ij} + C_{ij}$
    }\\
    \includegraphics{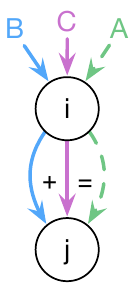}
    \subcaption{
      Iteration graph workspace optimizations
    }
    \label{fig:case-spadd-original}
  \end{minipage}
  \begin{minipage}{0.59\linewidth}
    \center
    {
      $\iteration{i}
        ( \iteration{j} A_{ij} \assign w_j) \, \where \;
        ( \iteration{j} w_j \assign B_{ij} \seq
         \iteration{j} w_j \reduce{+} C_{ij} )
      $
    }\\
    \includegraphics{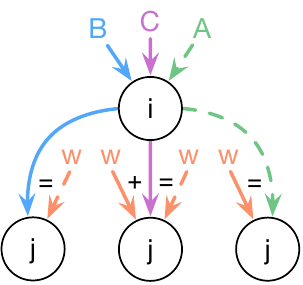}
    \subcaption{
      Iteration graph after workspace optimizations
    }
    \label{fig:case-spadd-split}
  \end{minipage}
  \caption{
    Sparse matrix addition $A_{ij} = B_{ij} + C_{ij}$. Splitting the addition
    and assignment operators removes expensive merge code at the cost of
    reduced temporal locality.  The code before and after the split are shown
    in in~\figref{examples-matadd}.
  }
  \label{fig:case-spadd}
\end{figure}

% Sparse matrix addition
Sparse matrix addition demonstrates the workspace optimization for addition
operators.  Sparse additions result in code to iterate over the union of the
nonzeros of the operands, as a multi-way merge with three loops
~\cite{knuth1973}.  \figref{case-spadd-original} shows the concrete index
notation and iteration graph for a sparse matrix addition.  When the matrices
are stored in the CSR format, which is sparse in the second dimension, the
compiler must emit code to merge $B$ and $C$ at the $j$ index variable.  Such
merge code contains many if statements that are expensive on modern processors.
Merge code also grows exponentially with the number of additions, so if many
matrices are added it is necessary to either split the input expression or,
better, to use the workspace optimization at the inner index variable so that
the outer loop can still be shared.

% Operator split
Applying the workspace optimization twice to both $B$ and $C$ at $j$ introduces
a dense row workspace that rows of $B$ and $C$ are in turn are added into, and
that is then copied over to $A$.  The resulting code was shown
in~\figref{examples-matadd} and has decreased temporal locality due to the
workspace reuse distance, but avoids expensive merges.  Whether this results in
an overall performance gain depends on the machine, the number of operands that
are merged, and the nonzero structure of the operands.  We show results
in~\figref{spadd_results}.

\subsection{Matricized Tensor Times Khatri-Rao Product}
\label{sec:case-mttkrp}

\begin{figure*}
  \begin{minipage}[b]{0.20\linewidth}
    \center
    \includegraphics{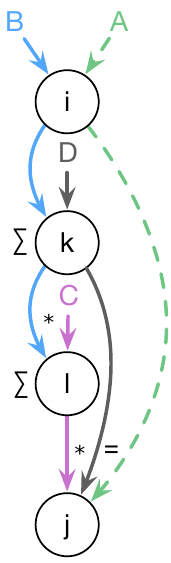}
    \vspace{21px}
    \subcaption{
      Before workspace optimization
    }
    \label{fig:case-mttkrp-original}
  \end{minipage}
  \vspace{10px}
  \hspace{0.01\linewidth}
  \begin{minipage}[b]{0.77\linewidth}
    \begin{minipage}{1\linewidth}
    \center
    \hspace{-10px}
    {\small
      $\iteration{ik}
        \left( \iteration{j} A_{ij} \reduce{+} w_{j} D_{kj} \right) \where
        \left( \iteration{lj} w_{j} \reduce{+} B_{ikl} C_{lj} \right)
      $
    }\\
    \end{minipage}
    \begin{minipage}[b]{0.24\linewidth}
      \flushleft
      \includegraphics{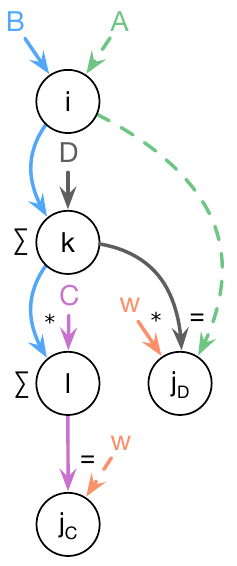}
      \vspace{20px}
    \end{minipage}
    \begin{minipage}[b]{0.75\linewidth}
      \flushright
      \vspace{5px}
      \includegraphics{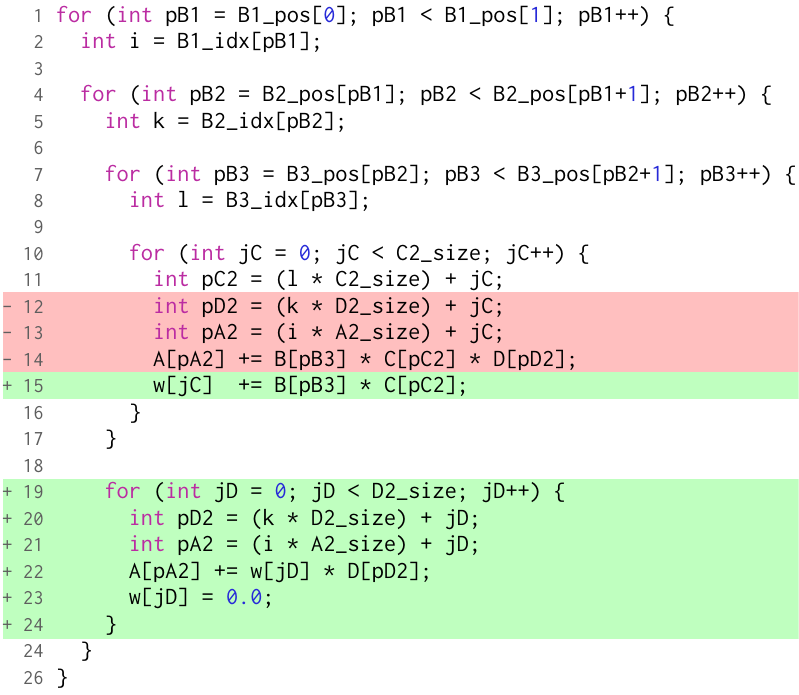}
    \end{minipage}
    \subcaption {\label{fig:case-mttkrp-split}\label{fig:code-mttkrp-split}
      After optimization to pre-compute $B_{ikl} C_{lj}$ in workspace $w$
      at $j$.  The code diff shows the effect of the transformation.
    }
  \end{minipage}
  \begin{minipage}{\linewidth}
    \center
    \vspace{2mm}
    {\small
      $\iteration{i}
        \left( \iteration{j} A_{ij} \assign v_{j} \right) \where
        \left( \iteration{k}
          \left( \iteration{j} v_{j} \reduce{+}  w_{j} D_{kj} \right) \where
          \left( \iteration{lj} w_{j} \reduce{+} B_{ikl} C_{lj} \right)
        \right)
      $
    }\\
    \vspace{-2mm}
  \vspace{3px}
  \begin{minipage}[b]{0.408\linewidth}
    \centering
    \includegraphics{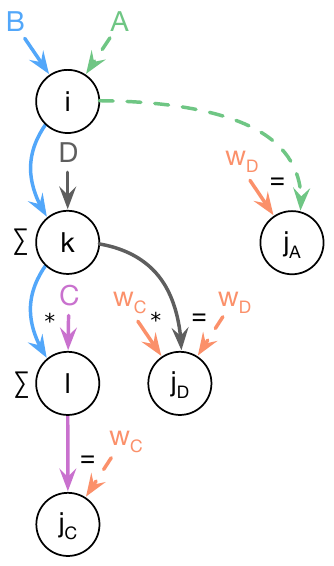}
    \vspace{20px}
  \end{minipage}
  \begin{minipage}[b]{0.582\linewidth}
    \centering
    \vspace{8px}
    \includegraphics{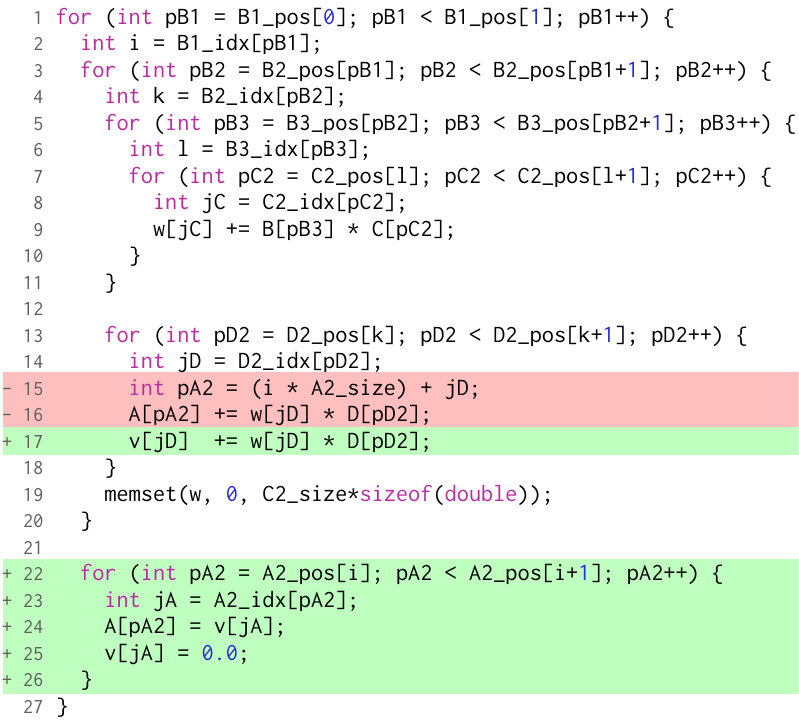}
  \end{minipage}
  \subcaption { \label{fig:case-mttkrp-splits} \label{fig:code-mttkrp-splits}
    After further optimization to pre-compute $w_j D_{kj}$ in workspace $v$ at
    $j$.  The code diff shows the effect of the transformation.
  }
  \end{minipage}
  \caption { \label{fig:case-mttkrp}
    The matricized tensor times Khatri-Rao product (MTTKRP) $A_{ij} = \sum_{kl}
    B_{ikl} C_{lj} D_{kj}$.  Workspacing $B_{ikl} C_{lj}$ at $j$ hoists the
    expression out of the $l$ loop and therefore removes redundant
    loop-invariant work.  If the matrix $A$ is sparse, then also workspacing
    $w_j D_{kj}$ at $j$ introduces a random access workspace that removes the
    need to insert into $A$.
  }
\end{figure*}

% MTTKRP
The matricized tensor times Khatri-Rao product (MTTKRP) is the critical kernel
in the alternating least squares algorithm to compute the canonical polyadic
decomposition of tensors~\cite{hitchcock1927}.  The canonical polyadic
decomposition generalizes the singular value decomposition to higher-order
tensors, and has applications in data analytics~\cite{cichocki2014}, machine
learning~\cite{phan2010}, neuroscience~\cite{mocks1988}, image classification
and compression~\cite{shashua2001}, and other fields~\cite{kolda2009}.

% Before optimization
The MTTKRP can be expressed with tensor index notation as \mbox{$A_{ij} =
\sum_{kl} B_{ikl} C_{lj} D_{kj}$}.  That is, we multiply a three-dimensional
tensor by two matrices in the $l$ and $k$ dimensions.  These simultaneous
multiplications require four nested loops.  \figref{case-mttkrp-original} shows
the iteration graph before optimization, where the matrices are stored
row-major.  The iteration graph results in four nested loops.  The three
outermost loops iterate over the sparse data structure of $B$, while the
innermost loop iterates over the range of the $j$ index variable.

After applying the workspace optimization to the expression $B_{ikl} C_{lj}$ at
$j$ we get the iteration graph in \figref{case-mttkrp-split}.  The index
variable $j$ has been split in two.  The second $j$ is no longer dominated by
$l$ and is therefore evaluated higher up in the resulting loop nest.
Furthermore, if the matrices $C$ and $D$ are sparse in the second dimension,
then the workspace optimization also removes the need to merge their sparse
data structures.  The code listing in \figref{code-mttkrp-split} shows a code
diff of the effect of the optimization on the code when the matrices are dense.
The code specific to the iteration graph before optimizing is colored red, and
the code specific to the iteration graph after optimizing is colored green.
Shared code is not colored.  The workspace optimization results in code where
the loop over $j$, that multiplies $B$ with $D$, has been lifted out of the $l$
loop, resulting in fewer total multiplication.  The drawback is that the
workspace reduces temporal locality, as the reuse distance between writing
values to it and reading them back can be large.  Our evaluation
in~\figref{mttkrp_results} shows that this optimization can result in
significant gains on large data sets.

The MTTKRP kernel does two simultaneous matrix multiplications.  Like the
sparse matrix multiplication kernel in~\secref{case-spmm}, it scatters values
into the middle of the result matrix $A$.  The reason is that the $j$ index
variables are dominated by reduction variables.  If the matrix $A$ is sparse
then inserts are expensive and the code profits from applying the workspace
optimization again to pre-compute $w_j D_{kj}$ in a workspace, as shown
in~\figref{case-mttkrp-splits}.  The effect is that values are scattered into a
dense workspace with random access and copied to the result after a full row of
the result has been computed.  \figref{code-mttkrp-splits} shows a code diff of
the effect of making the result matrix $A$ sparse and pre-computing $w_j
D_{kj}$ in a workspace $v$.  Both the code from before optimization (red) and
the code after (green) assumes the operand matrices $C$ and $D$ are sparse, as
opposed to \figref{code-mttkrp-split} where $C$ and $D$ were dense.  As in the
sparse matrix multiplication code, the code after the workspace optimization
scatters into a dense workspace $v$ and, when a full row has been computed,
appends the workspace nonzeros to the result.

\section{Evaluation}
\label{sec:evaluation}

In this section, we evaluate the effectiveness of the workspace optimization by
comparing the performance of sparse kernels with workspaces against
hand-written state-of-the-art sparse libraries for linear and tensor algebra.

\begin{table}
  \caption{\label{tab:test_data}
    Test matrices from the SuiteSparse Matrix Collection~\cite{suitesparse} and
    test tensors from the FROSTT Tensor Collection~\cite{frostt}.
  }
  \small
  \begin{tabular}{llrr}
    \hline
    \textbf{Tensor} & \textbf{Domain} & \textbf{NNZ} & \textbf{Density}\\
    \hline
    bcsstk17 & Structural & 428,650 & 4E-3\\
    pdb1HYS  & Protein data base & 4,344,765 & 3E-3\\
    rma10    & 3D CFD & 2,329,092 & 1E-3\\
    cant     & FEM/Cantilever & 4,007,383 &  1E-3\\
    consph   & FEM/Spheres & 6,010,480 & 9E-4\\
    cop20k   & FEM/Accelerator & 2,624,331 & 2E-4\\
    shipsec1 & FEM & 3,568,176 & 2E-4\\
    scircuit & Circuit & 958,936 & 3E-5\\
    mac-econ & Economics & 1,273,389 & 9E-5\\
    pwtk     & Wind tunnel & 11,524,432 & 2E-4\\
    webbase-1M & Web connectivity & 3,105,536 & 3E-6\\\hline
    Facebook & Social Media & 737,934 & 1E-7\\
    NELL-2   & Machine learning & 76,879,419 & 2E-5\\
    NELL-1   & Machine learning & 143,599,552 & 9E-13\\\hline
  \end{tabular}
\end{table}

\subsection{Methodology}

All experiments are run on a dual-socket 2.5 GHz Intel Xeon
E5-2680v3 machine with 12 cores/24 threads and 30 MB of L3 cache per socket,
running Ubuntu 14.04.5 LTS.  The machine contains 128~GB of memory and runs
Linux kernel version 3.13.0 and GCC 5.4.0.  For all experiments, we ensure the
machine is otherwise idle and report average cold cache performance, without
counting the first run, which often incurs dynamic loading costs and other
first-run overheads.  The experiments are single-threaded unless otherwise
noted.

We evaluate our approach by comparing performance on linear algebra kernels
with Eigen~\cite{eigen} and Intel MKL~\cite{mkl} 2018.0, and  tensor algebra
kernels against the high-performance SPLATT library for sparse tensor
factorizations~\cite{smith2015}.  We obtained real-world matrices and tensors
for the experiments in Sections~\ref{sec:spmm} and~\ref{sec:mttkrp} from the
SuiteSparse Matrix Collection~\cite{suitesparse} and the FROSTT Tensor
Collection~\cite{frostt} respectively.  Details of the matrices and tensors
used in the experiments are shown in Table~\ref{tab:test_data}.  We constructed
the synthetic sparse inputs using the random matrix generator in \tool{}, which
places nonzeros randomly to reach a target sparsity.  All sparse matrices are
stored in the compressed sparse row (CSR) format.

\subsection{Sparse Matrix-Matrix Multiplication}
\label{sec:spmm}

Fast sparse matrix multiplication (SpMM) algorithms use workspaces to store
intermediate values~\cite{gustavson1978}.  We compare our generated workspace
algorithm to the SpMM implementations in MKL and Eigen.   We compute SpMM
with two operands: a real-world matrix from Table~\ref{tab:test_data} and a
synthetic matrix generated with a specific target sparsity, with uniform random
placement of nonzeros.  Eigen implements a \textit{sorted} algorithm, which
sorts the column entries within each row so they are ordered, while MKL's
\code{mkl_sparse_spmm} implements an \textit{unsorted} algorithm---the column
entries may appear in any order.\footnote{According to MKL documentation, its
sorted algorithms are deprecated and should not be used.}  Because these two
algorithms have very different costs, we compare to a workspace variant of
each.  In addition, we evaluate two variants of workspace algorithm: one that
separates assembly and computation, and one that fuses the two operations.  The
approach described by~\citeauthor{kjolstad2017} can in theory handle sparse
matrix multiplication by inserting into sparse results.  The current
implementation\footnote{As of Git revision bf68b6.}, however, does not support
this, so we do not compare against it.

\begin{figure}
\centering
  \includegraphics[width=\columnwidth]{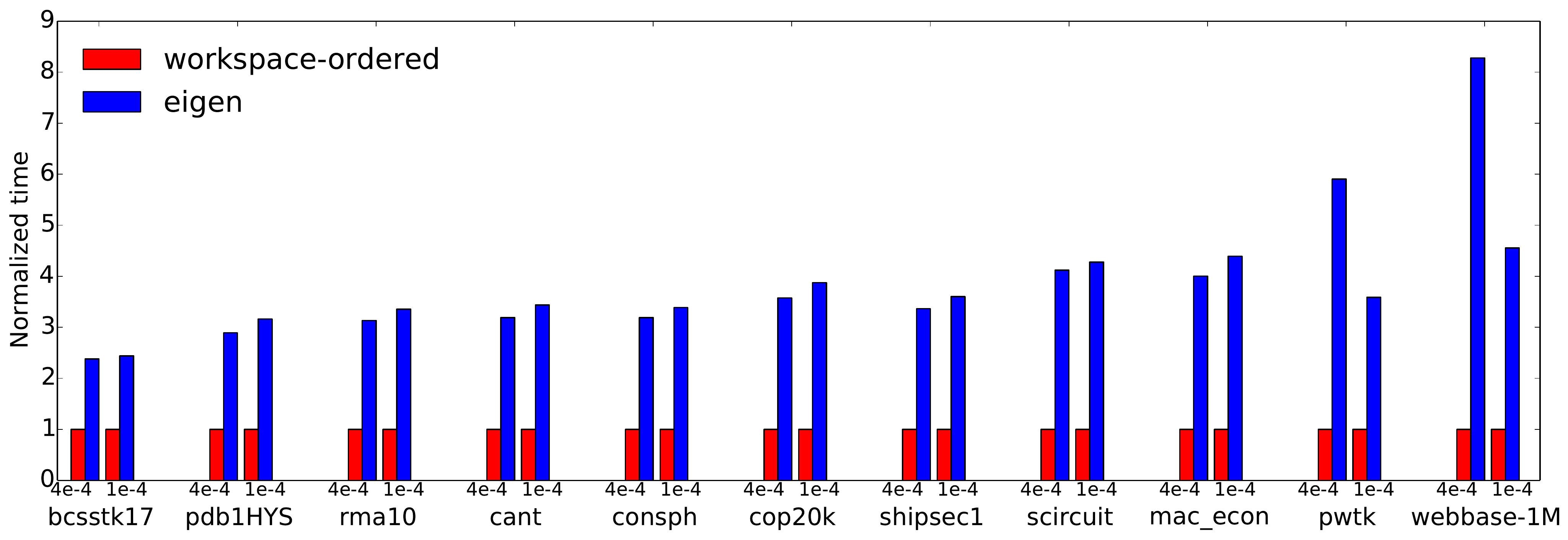}
  \includegraphics[width=\columnwidth]{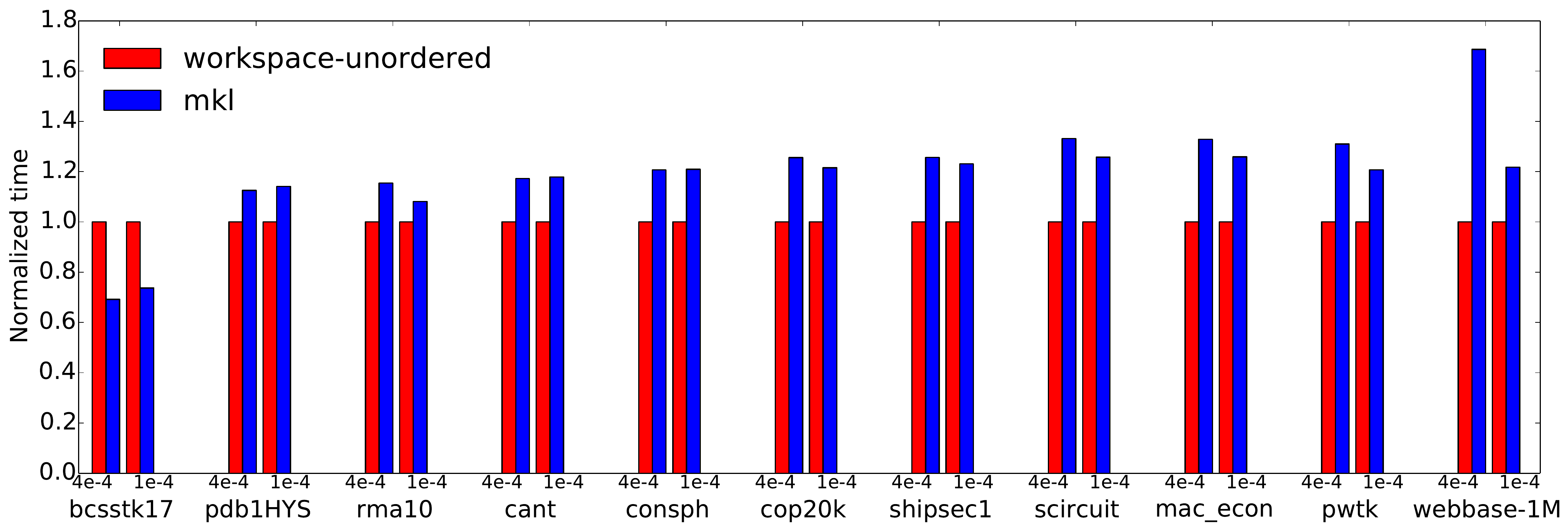}
  \caption{\label{fig:spmm_results} Sparse matrix multiplication results for
  the matrices in Table~\ref{tab:test_data}.  We show relative runtime for both
  sorted (top) and unsorted column entries (bottom); Eigen's algorithm sorts them while MKL's
  \code{mkl_sparse_spmm} function leaves them unsorted.}
\end{figure}

\newcolumntype{R}[2]{%
  >{\adjustbox{angle=#1,lap=\width-(#2)}\bgroup}%
  l%
  <{\egroup}%
}
\newcommand*\rot{\multicolumn{1}{R{45}{1em}}}% no optional argument here, please!

\begin{table}
  \small
  \setlength{\tabcolsep}{3.7pt}
  \caption{\label{tab:spmm_breakdown}
    Breakdown of time, in milliseconds (with 4 significant digits), to multiply
    the test matrices in Table~\ref{tab:test_data} with a random operand of
    density 4E-4.  Running time is given separately for the workspace assemble
    and compute kernels, as well as the variant that assembles and computes in one
    kernel (fused).  Times are compared to the total time spent by Eigen and
    MKL. For MKL, we use
    \code{mkl_sparse_spmm}, which does not sort rows of the output matrix.
  }
  \begin{tabular}{l r r r r r r r r r r r}
    & \rot{bcsstk17} & \rot{bcsstk17} & \rot{rma10} & \rot{cant} & \rot{consph} & \rot{cop20k} & \rot{shipsec1} & \rot{scircuit} & \rot{mac-econ} & \rot{pwtk} & \rot{webbase-1M}\\\hline
    \multicolumn{12}{c}{\textbf{Sorted (ms)}}\\\hline
assembly & 47.04 & 1867 & 1223 & 2937 & 6021 & 3445 & 12700 & 1691 & 2642 & 29930 & 37670\\
compute & 6.703 & 373.1 & 276.1 & 655.8 & 1397 & 937.5 & 3322 & 525.5 & 846.0 & 	8229 & 11000\\
assembly+compute & 53.74 & 2241 & 1499 & 3593 & 7418 & 4383 & 16020 & 2217 & 3489 & 38160 & 48670\\
fused & 51.18 & 2099 & 1397 & 3328 & 6841 & 3920 & 14800 & 2025 & 3207 & 35350 & 43720\\
Eigen & 121.7 & 6068 & 4378 & 10620 & 21820 & 14020 & 49840 & 8342 & 12840 & 208700 & 361900\\\hline
    \multicolumn{12}{c}{\textbf{Unsorted (ms)}}\\\hline
assembly & 5.469 & 209.6 & 153 & 355.4 & 723.1 & 461.9 & 1579 & 241.1 & 388.1 & 4123 & 6046\\
compute & 7.074 & 396.3 & 277.7 & 651.1 & 1402 & 960.1 & 3349 & 527.1 & 846.3 & 8295 & 9953\\
assembly+compute & 12.54 & 605.9 & 430.7 & 1006 & 2125 & 1422 & 4929 & 768.1 & 1234 & 12420 & 16000\\
fused & 12.1 & 464.3 & 325.7 & 752.5 & 1610 & 1081 & 3859 & 578.9 & 951 & 9454 & 11320\\
MKL & 8.371 & 522.7 & 375.9 & 882.1 & 1943 & 1357 & 4847 & 770.7 & 1264 & 12380 & 19090\\\hline
  \end{tabular}
\end{table}

Figure~\ref{fig:spmm_results} shows running times for sparse matrix
multiplication for each matrix in Table~\ref{tab:test_data} multiplied by a
synthetic matrix of nonzero densities 1E-4 and 4E-4, using our fused workspace
implementation.  On average, Eigen is slower than our approach, which generates
a variant of Gustavson's matrix multiplication algorithm, by 4$\times$ and
3.6$\times$ respectively for the two sparsity levels.  For the unsorted
algorithm, we compare against Intel MKL, and find that our performance is 28\%
faster and 16\% on average.  The generated workspace algorithm is faster (by up
to 68\%) than MKL's hand-optimized SpMM implementation in all but one case,
which is 31\% slower.

Table~\ref{tab:spmm_breakdown} breaks down the running times for the different
codes for multiplying with a matrix of density 4E-4.  Due to sorting, assembly
times for the sorted algorithm are quite large; however, the compute time is
occasionally faster than the unsorted compute time, due to improved locality
when accumulating workspace entries into the result matrix.  The fused
algorithm is also faster when not using sorting, because otherwise the sort
dominates the time (we use the standard C \code{qsort}).

\subsection{Matricized Tensor Times Khatri-Rao Product}
\label{sec:mttkrp}

\begin{figure*}[tb]
\centering
  \includegraphics[width=0.5\columnwidth]{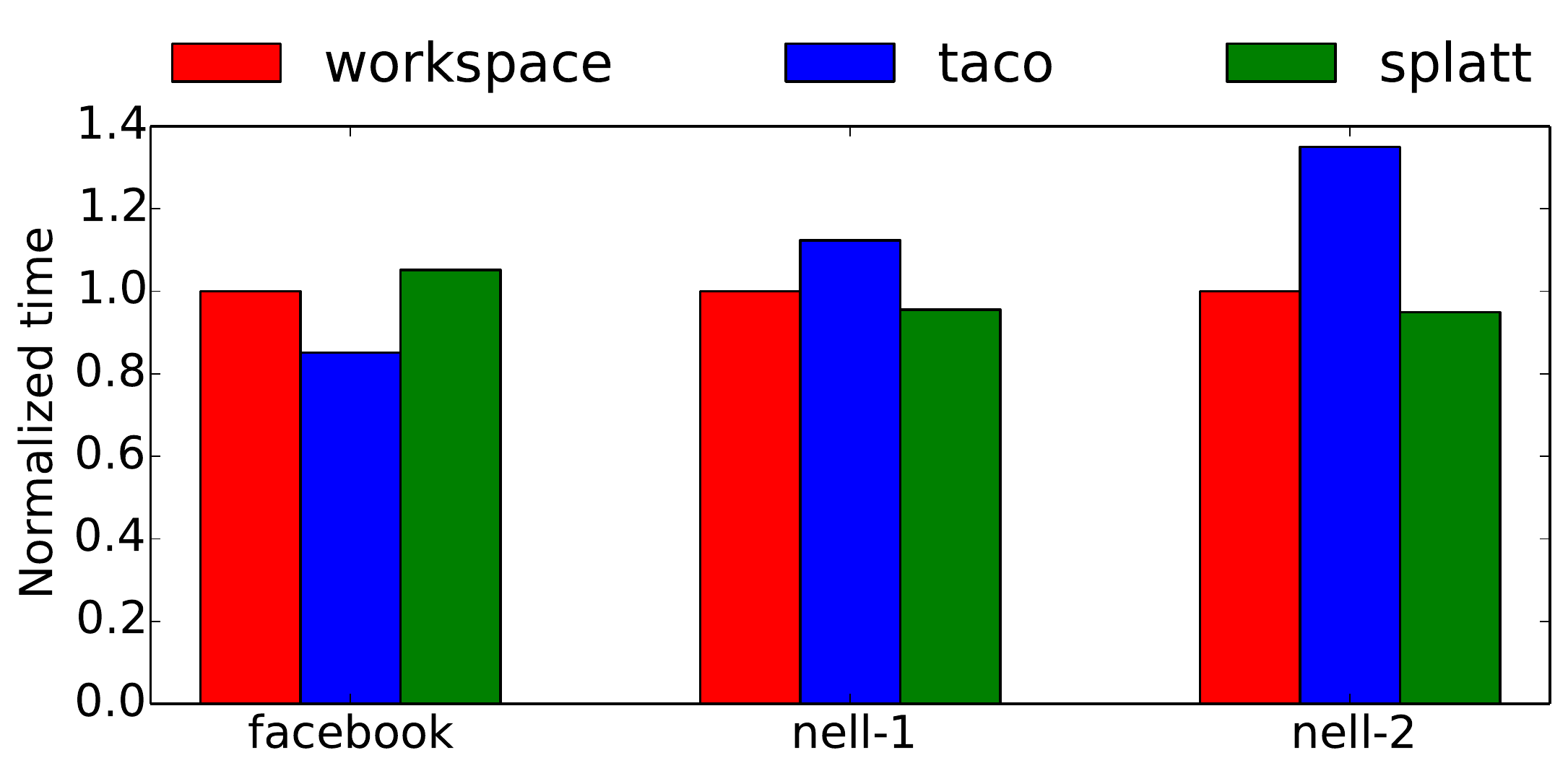}
  \caption{\label{fig:mttkrp_results}
    Matricized tensor times Khatri-Rao product (MTTKRP) running times,
    normalized to the workspace algorithm running time.  MTTKRP is run in parallel
    using \texttt{numactl} to restrict execution to a single socket.  Only compute times are
    shown; assembly times are negligible because the outputs are dense.
  }
\end{figure*}

Matricized tensor times Khatri-Rao product (MTTKRP) is used to compute
generalizations of SVD factorization for tensors in data analytics.  The
three-dimensional version takes as input a sparse 3-tensor and two matrices,
and outputs a matrix.  Figure~\ref{fig:mttkrp_results} shows the results for
our workspace algorithm on three input tensors, compared to \tool{} and the
hand-coded SPLATT library.  We show only compute times, as the assembly times
are negligible because the outputs are dense. We compare parallel single-socket
implementations, using \texttt{numactl} to restrict execution to a single
socket.

For the NELL-1 and NELL-2 tensors, the workspace algorithm outperforms the
merge-based algorithm in \tool{} and is within 5\% of the hand-coded
performance of SPLATT.  On the smaller Facebook dataset, the merge algorithm is
faster than both our implementation and SPLATT's.  That is, different inputs
perform better with different algorithms, which demonstrates the advantage of
being able to generate both versions of the algorithm.

\subsection{Matricized Tensor Times Khatri-Rao Product with Sparse Matrices}
\label{sec:spmttkrp}

It is useful to support MTTKRP where both the tensor and matrix operands are
sparse~\cite{smith2017}.  If the result is also sparse, then the MTTKRP can be
much faster since it only needs to iterate over nonzeros.  The code is tricky
to write, however, and cannot be generated by the current version of \tool{},
although the prior merge-based theory supports it.   In this section, we use a
workspace implementation of sparse MTTKRP enabled by the workspace
optimization.  As far as we are aware, ours is the first implementation of an
MTTKRP algorithm where all operands are sparse and the output is a sparse
matrix.  Because we have not implemented a parallel version of MTTKRP with sparse
outputs, we perform this comparison with single-threaded implementations of both
MTTKRP versions.
\begin{figure*}
  \centering
  \begin{minipage}{0.3\textwidth}
    \centering
    \includegraphics[width=\textwidth]{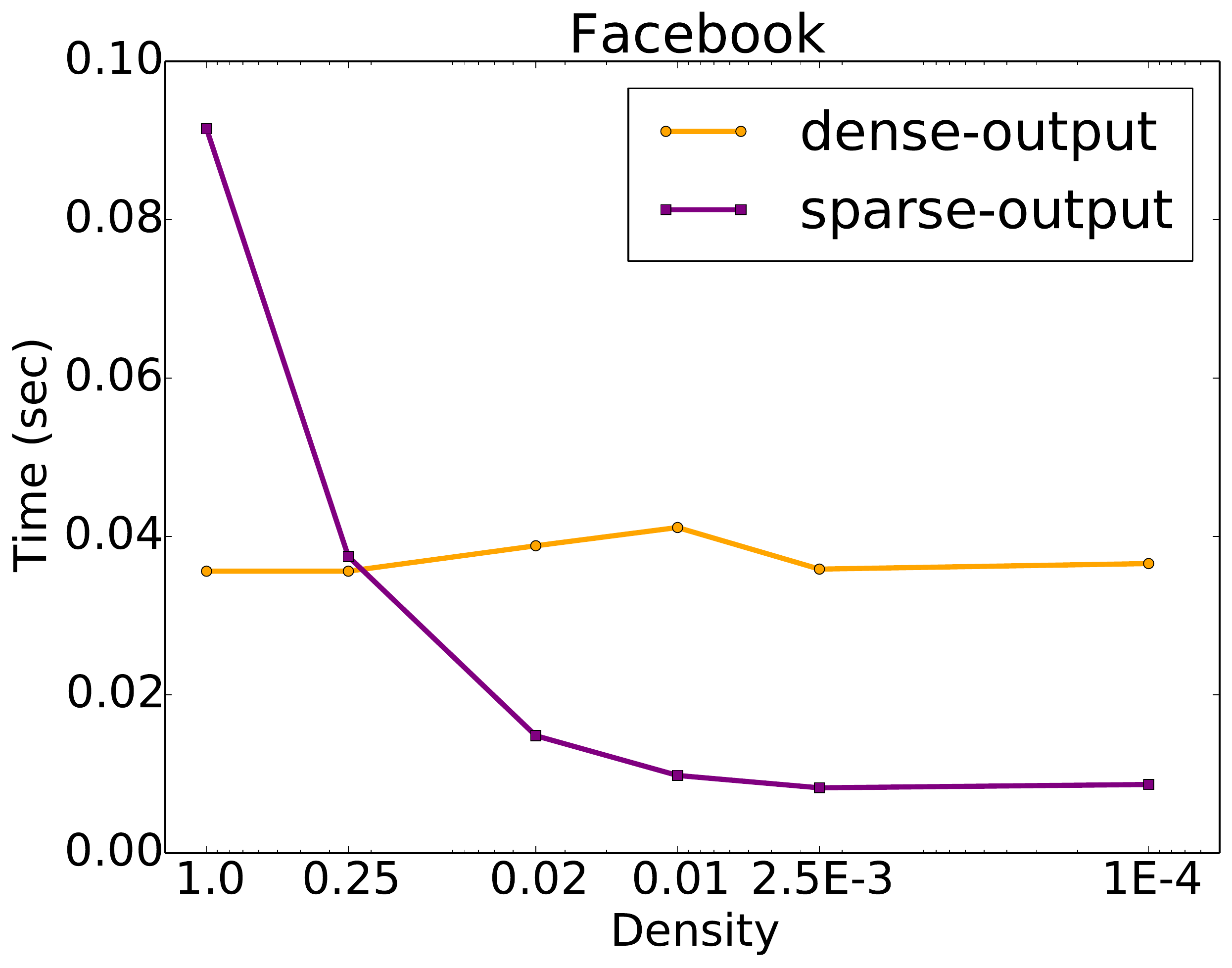}
  \end{minipage}
  \begin{minipage}{0.3\textwidth}
    \centering
    \includegraphics[width=\textwidth]{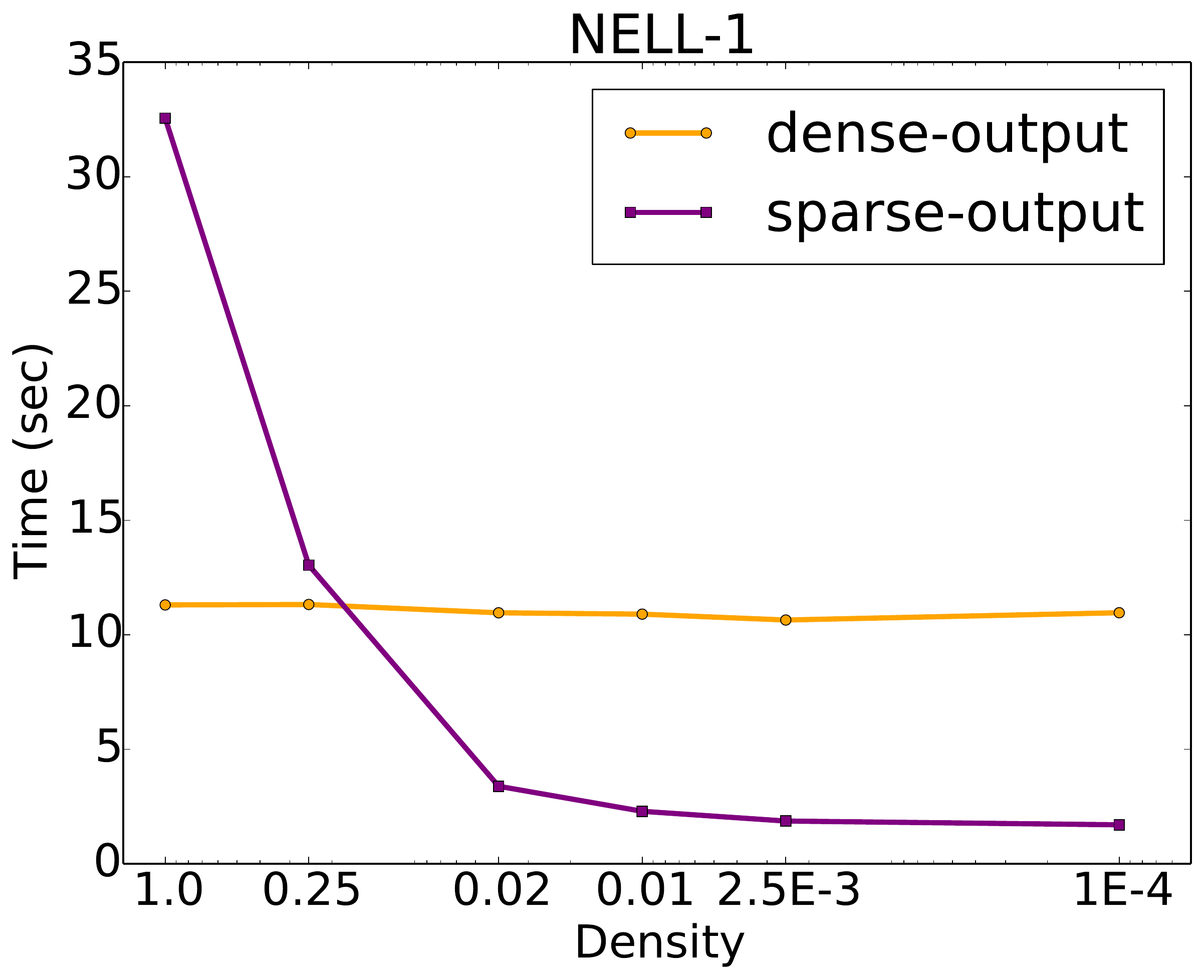}
  \end{minipage}
  \begin{minipage}{0.3\textwidth}
    \centering
    \includegraphics[width=\textwidth]{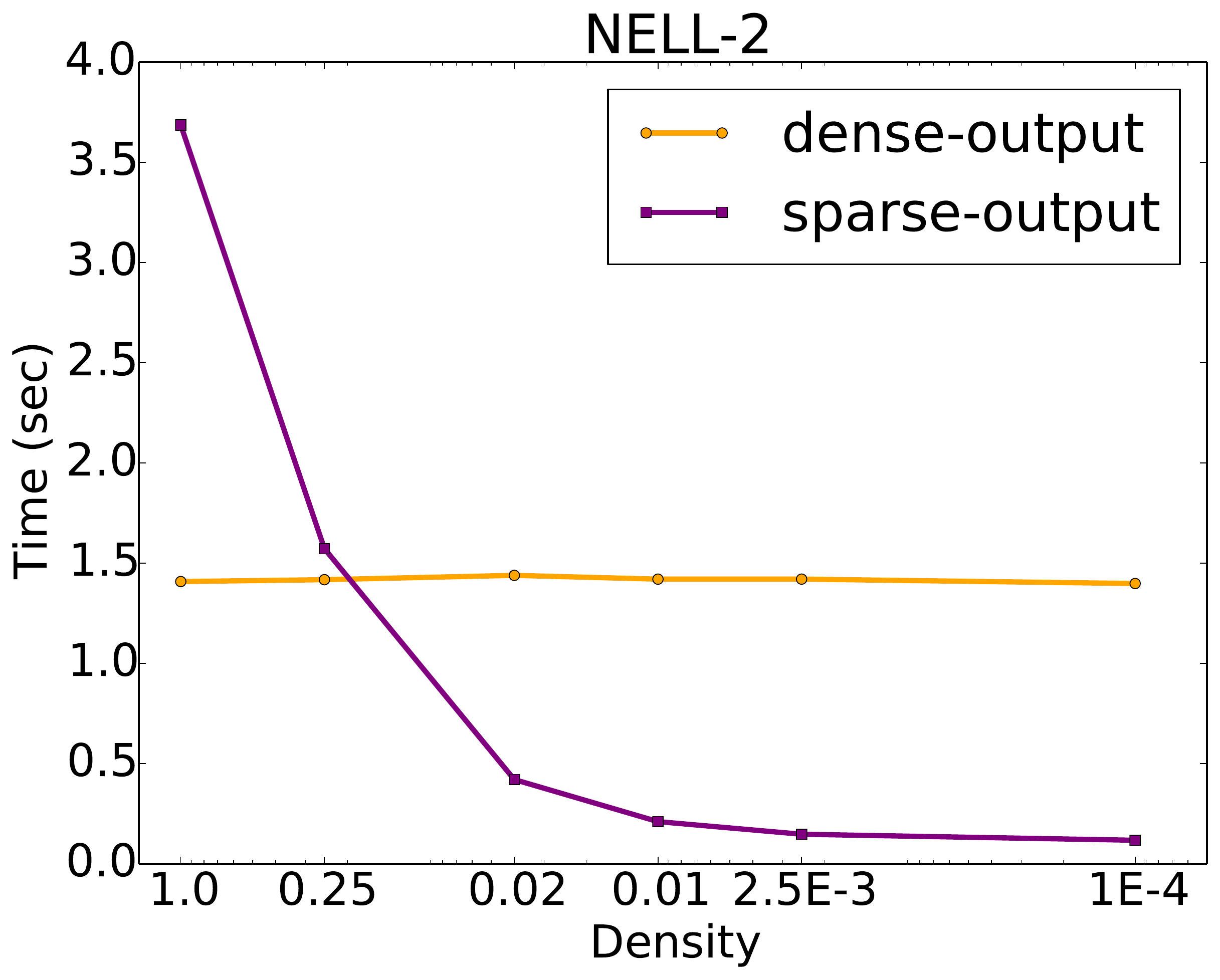}
  \end{minipage}
  \caption{
    MTTKRP compute time as we vary the density of the matrix operands, for the
    three test tensors.  We compare MTTKRP computed with a workspace when the
    matrix operands are passed in as dense matrices with a dense output against an implementation
    that takes sparse matrices as inputs and outputs a sparse matrix.  In all
    cases, the tensor is passed in using a sparse format.  This comparison uses
    single-threaded performance, as we have not implemented a parallel MTTKRP
    with sparse output.
  }
  \label{fig:spmttkrp-results}
\end{figure*}

Which version is faster depends on the density of the sparse operands.
Figure~\ref{fig:spmttkrp-results} shows experiments that compares the compute
times for MTTKRP with sparse matrices against MTTKRP with dense matrices, as we
vary the density of the randomly generated input matrices.  Note that the dense
matrix version should have the same performance regardless of sparsity and any
variation is likely due to system noise.  For each of the tensors, the
crossover point is at about 25\% nonzero values, showing that such a sparse
algorithm can be faster even with only a modest amount of sparsity in the
inputs. At the extreme, matrix operands with density 1E-4 can obtain speedups
of 4.5--11$\times$ for our three test tensors.

\begin{figure}
  \centering
  \begin{minipage}[c]{0.5\linewidth}
  \includegraphics[width=\linewidth]{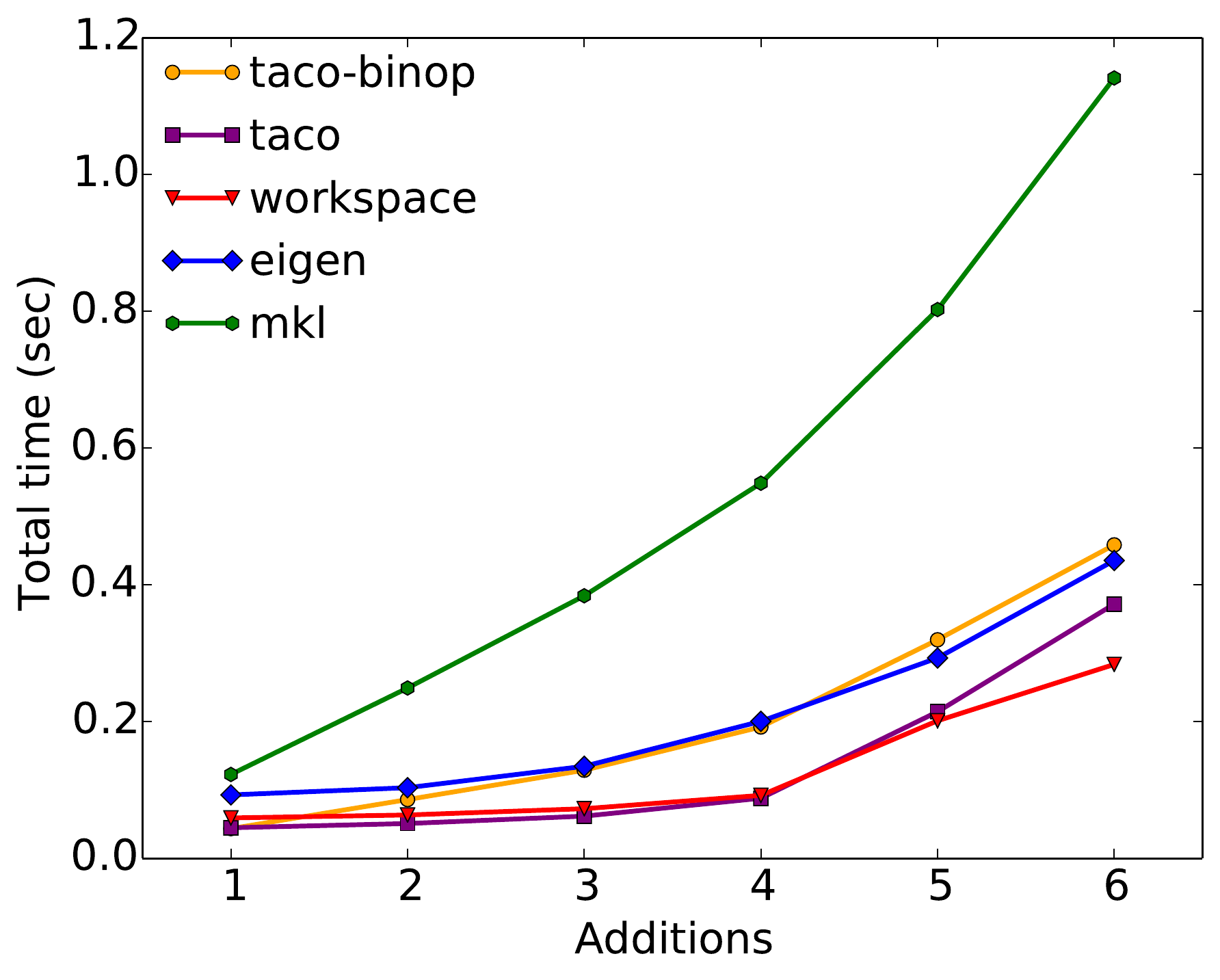}
  \end{minipage}\hspace*{10pt}%
  \begin{minipage}[b]{0.4\linewidth}
  \begin{tabular}{lrr}
    \textbf{Code} & \textbf{Assembly} & \textbf{Compute}\\\hline
    taco binop & 247 & 211\\
    taco & 190 & 182\\
    workspace & 190 & 93.3\\
    Eigen & & 436\\
    MKL & & 1141\\\hline
  \end{tabular}
  \end{minipage}
  \caption {\label{fig:spadd_results}%
  \label{fig:evaluation-matrix-add-scaling}
    Left: Scaling plot showing the time to assemble and compute $n$ matrix additions with Eigen,
    MKL, taco binary operations, a single multi-operand taco function, and workspaces.
    The matrices are described in Table~\ref{tab:spadd_breakdown}.
    Right: \label{tab:spadd_breakdown} Breakdown of sparse matrix addition time
    in ms for adding 7 matrices, for all codes
    The operands are randomly-generated sparse matrices of density
    2.56E-02,
    1.68E-03,
    2.89E-04,
    2.50E-03,
    2.92E-03,
    2.96E-02,
    1.06E-02,
    respectively.
  }
\end{figure}
\subsection{Sparse Matrix Addition}

To demonstrate the utility of workspaces for sparse matrix addition (SpAdd), we
show that the algorithm scales as we increase the number of operands.  In
Figure~\ref{fig:spadd_results}, we compare the workspace algorithm to \tool{}
using binary operations (as a library would be implemented), \tool{} generating
a single function for the additions, Intel MKL (using its inspector-executor
SpAdd implementation), and Eigen.  We pre-generate $k$ matrices with the target
sparsities chosen uniformly randomly from the range [1E-4, 0.01] and always add in
the same order and with the same matrices for each library.

The results of this experiment show two things.  First, that the libraries are
hampered by the restriction that they perform addition two operands at a time,
having to construct and compute multiple temporaries, resulting in less
performance than is possible using code generation.  Even given this approach,
\tool{} is faster than Intel MKL by 2.8$\times$ on average, while Eigen and
\tool{} show competitive performance.

Secondly, the experiment shows the value of being able to produce both
merge-based and workspace-based implementations of SpAdd.  At up to four
additions, the two versions are competitive, with the merge-based code being
slightly faster.  However, with increasing numbers of additions, the workspace
code begins to outperform the \tool{} implementation, showing an increasing gap
as more operands are added.  Table~\ref{tab:spadd_breakdown} breaks down the
performance of adding 7 operands, separating out assembly time for the
\tool{}-based and workspace implementations.  For this experiment, we reuse the
matrix assembly code produced by taco to assemble the output, but compute
using a workspace.  Most of the time is spent in assembly, which is unsurprising,
given that assembly requires memory allocations, while the computation performs only
point-wise work without the kinds of reductions found in MTTKRP and SpMM.

\section{Related Work}
\label{sec:related-work}

Related work is divided into work on tensor algebra compilation, work on manual
workspace optimizations of matrix and tensor kernels, and work on general loop
optimization.

% Tensor Algebra Compilation
There has been much work on optimizing dense matrix and tensor
computations~\cite{apl,wolfe1982,mckinley1996,tce}.  Researchers have also
worked on compilation and code generation of sparse matrix computations,
starting with the work of \citeauthor{dutch1}~\cite{dutch1}, the Bernoulli
system~\cite{bernoulli}, and SIPR~\cite{sipr}.  Recently,
\citeauthor{kjolstad2017}~\shortcite{kjolstad2017} proposed a tensor algebra
compilation theory that compiles tensor index notation on dense and sparse
tensors.  These sparse compilation approaches, however, did not generate sparse
code with tensor workspaces to improve performance.

% General loop optimizations
One use of the workspace optimization in loop nests, in addition to removing
multi-way merge code and scatters into sparse results, is to split apart
computation that may take place at different loop levels.  This results in
operations being hoisted to a higher loop nest.  Loop invariant code motion has
a long history in compilers, going back to the first FORTRAN compiler in
1957~\cite{backus1978}.  Recently, researchers have found new opportunities for
removing redundancy in loops by taking advantage of high-level algebraic
knowledge~\cite{ding2017}.  Our workspace optimization applies to sparse tensor
algebra and can remove loop reduncancies from sparse code with indirect-access
loop bounds and many conditional branches.

% General sparse loop optimizations
The polyhedral model was originally designed to optimize dense loop nests
with affine loop bounds and affine accesses into dense arrays.  Sparse code,
however, involves nested indirect array accesses.  Recent work has to extend
the polyhedral model to these situations~\cite{strout2012, belaoucha2010,
chill, venkat2016automating}, using a combination of compile-time and runtime techniques,
but the space of loop nests on nested indirect array accesses is
complicated, and it difficult for compilers to determine when linear-algebraic
optimizations are applicable to the operations that the code represents.  Our
workspace optimization applies to sparse tensor algebra at the concrete index
notation level, before sparse code is generated, which makes it possible to
perform aggressive optimizations and convenient to reason about legality.

% Manual workspace optimizations
The first use of dense workspaces for sparse matrix computations is
\citeauthor{gustavson1978}'s sparse matrix multiplication implementation, that
we recreate with the workspace optimization in~\figref{case-spmm} to produce
the code in and~\figref{matmul-workspace}~\cite{gustavson1978}.  A workspace
used for accumulating temporary values is referred to as an expanded real
accumulator in~\cite{pissanetzky1984} and as an abstract sparse accumulator
data structure in~\cite{gilbert1992}.  Dense workspaces and blocking are used
to produce fast parallel code by~\citeauthor{patwary2015}~\cite{patwary2015}.
They also tried a hash map workspace, but report that it did not have good
performance for their use.  Furthermore, \citeauthor{bulucc2009} use blocking
and workspaces to develop sparse matrix-vector multiplication algorithms for
the CSB data structure that are equally fast for $Ax$ and
$A^Tx$~\cite{bulucc2009}.  Finally, \citeauthor{smith2015} uses a workspace to
hoist loop-invariant code in their implementation of MTTKRP in the SPLATT
library~\cite{smith2015}.  We re-create this optimization with the workspace
optimization in~\figref{case-mttkrp-split} and show the resulting source code
in~\figref{code-mttkrp-split}.

\section{Conclusion}
\label{sec:conclusions}

This paper presented the concrete index notation optimization language for
describing how tensor index notation should execute and a workspace
optimization that introduces workspaces to remove insertion into sparse
results, conditionals, and to hoist loop-invariant computations.  The
optimization enables a new class of sparse tensor computations with sparse
results and improves performance of other tensor computations to match
state-of-the-art hand-optimized implementations.  We believe the importance of
workspaces will increase in the future as combining new tensor formats will
require workspaces as glue.  Furthermore, we believe the concrete index
notation language can grow into a language for general tensor optimization,
including loop tiling, strip-mining, and splitting.  Combined with a scheduling
language to command these concrete index notation transformations, the
resulting system separates algorithm from schedule.  This lets end users
specify the computation they want, in tensor index notation, while the
specification for how it should execute can be specified by performance
experts, autotuning systems, machine learning, or heuristics.

\bibliography{paper} 

\end{document}